\documentclass[10pt]{article}
\usepackage[utf8x]{inputenc}
\usepackage{dsfont}
\begin{document}

\begin{center}
{\LARGE  The Standard Model with one universal extra dimension}
\end{center}

\vspace{0.03cm}

\begin{changemargin}{0.6in}{0.6in}

\begin{center}
A. Cordero--Cid$^{(a)}$, M. G\' omez--Bock$^{(b)}$, H. Novales--S\'anchez$^{(b)}$, and J. J. Toscano$^{(b)}$
\end{center}

\begin{center}
{\small $^{(a)}${\it Facultad de Ciencias de la Electr\'onica, Benem\'erita Universidad Aut\'onoma de Puebla, Blvd. 18 Sur y Av. San Claudio, 72590, Puebla, Pue., M\'exico.}
\\
$^{(b)}${\it Facultad de Ciencias F\'isico Matem\'aticas, Benem\'erita Universidad Aut\'onoma de Puebla, Apartado Postal 1152, Puebla, Puebla, M\'exico.}}
\end{center}

{\small Effects of universal extra dimensions on Standard Model observables first arise at the one--loop level. The quantization of this class of theories is therefore essential in order to perform predictions. A comprehensive study of the ${\rm SU_C}(3)\times{\rm SU_L}(2)\times{\rm U_Y}(1)$ Standard Model defined in a space--time manifold with one universal extra dimension, compactified on the oribifold $S^1/Z_2$, is presented. The fact that the four--dimensional Kaluza--Klein theory is subject to two types of gauge transformations is stressed and its quantization under the basis of the BRST symmetry discussed. A ${\rm SU_C}(3)\times {\rm SU_L}(2)\times {\rm U_Y}(1)$--covariant gauge--fixing procedure for the Kaluza--Klein excitations is introduced. The connection between gauge and mass eigenstate fields is established in an exact way. An exhaustive list of the explicit expressions for all physical couplings induced by the Yang--Mills, Currents, Higgs, and Yukawa sectors is presented. The one--loop renormalizability of the standard Green's functions, which implies that the Standard Model observables do not depend on a cutoff scale, is stressed.}
\end{changemargin}

\section{Introduction}
\label{in}
The phenomenological implications of extra dimensions on Standard Model (SM) observables have been the subject of considerable interest in the literature since Antoniadis, Arkani-Hamed, Dimopoulos, and Dvali~\cite{ED} argued that relatively large extra dimensions may be detected at the TeV scale. In most scenarios, our observed three-dimensional space is a 3-brane that is embedded in a higher $D$--dimensional spacetime, which is known as the bulk. If the additional dimensions are small enough, the SM gauge and matter fields are phenomenologically allowed to propagate in the bulk. Of course, if there are extra dimensions, they must be smaller than the smallest scale which has been currently explored by experiments. In this work, we will focus on a generalization of the SM to five dimensions in the universal extra dimensional (UED) context, and we will assume that the fifth dimension is compactified on the $S^1/Z_2$ orbifold with radius $R$. As a result of the compactification, each of the fields that propagate in the bulk expands into a series of states known as a Kaluza-Klein (KK) tower, with the individual KK excitations being labeled by mode numbers and the SM fields corresponding to the zero modes. It is well known that gauge theories in more than four dimensions are not renormalizable in the Dyson's sense, so that they must be recognized as effective theories that are embedded within some other consistent UV completion, such as string theories. The nonrenormalizable nature of higher--dimensional theories arises from the fact that their coupling constants are dimensionful. Although at the level of the four--dimensional theory the coupling constants are dimensionless and the corresponding Lagrangian does not involve interactions of canonical dimension higher than four, the nonrenormalizable character manifests itself through the infinite multiplicity of the KK modes. So, the effective theory must be cut off at some scale $M_{\rm s}$, above which the fundamental theory enters. In a recent paper~\cite{NT}, we examined the gauge structure and quantization of the four--dimensional KK theory that arises from a five--dimensional pure Yang--Mills theory, after dimensional reduction. In particular, we showed that this theory is subject to satisfy two types of gauge transformations, namely, the standard gauge transformations, which are obeyed by the zero mode gauge fields $A^{(0)a}_\mu$, and another sort of local gauge transformations, which we called  nonstandard gauge transformations, to which are subject the KK excitations $A^{(n)a}_\mu$~\cite{NT}. In that paper, we also showed that the SM, or light, Green's  functions (Green's functions whose external legs are all zero KK modes or, equivalently, SM fields) are renormalizable at the one--loop level. More recently, this fact was proven explicitly through the direct integration of the heavy KK excitations~\cite{NT2}. The cutoff insensitivity of light Green's functions at the one--loop level, which seems to be exclusive of UED models with one extra dimension, has already been pointed out in previous studies on some electroweak observables~\cite{PS,ACD} and verified very recently~\cite{FMNRT} for the case of one--loop radiative corrections to the trilinear $WW\gamma$ and $WWZ$ vertices. One peculiarity of UED models is that the tree--level couplings among KK excited modes and zero modes involves strictly two or more KK excitations. This means that the electroweak observables are insensitive to virtual effects of KK excitations at tree level, although they can receive contributions at the one--loop level or higher orders. The main goal of this work is to present a comprehensive study of the vertices involved in the theory, for we think that this important predictive power of UED theories in five dimensions deserves especial attention. This theory would serve as a basis to estimate in an unambiguous way the impact of extra dimensions on electroweak observables. We will present a detailed study of the tree--level structure of the four--dimensional KK theory. Our results comprise a complete list of the Lagrangians characterizing the vertices generated by the compactified theory, including the definition of a gauge--fixing procedure for both the SGT and the NSGT.

The rest of the paper has been organized as follows. In Sec. \ref{model}, the structure of the five--dimensional SM and the compactification of the extra dimension, including a gauge--fixing procedure, are discussed, whereas, in Sec. \ref{vt}, a detailed list of the physical vertices of the theory is presented. In Sec. \ref{c} the conclusions are presented. Finally, some technical details are presented in appendices.

\section{The Standard Model in five dimensions}
\label{model}
Consider the SM defined in a five--dimensional flat space--time, in which the fifth coordinate is compactified as a circle of radius $R$. From here on, the standard four--dimensional coordinates will be denoted by $x$, whereas the fifth--dimension coordinate will be represented by $y$. In the context of UED, one assumes that all fields and gauge parameters\footnote{Some phenomenological implications of having gauge parameters confined to the 3--brane are studied in~\cite{NT3}.} are periodic functions on this coordinate and expands them in Fourier series along it. In general, for a given field or gauge parameter, one has:
\begin{equation}
\varphi(x,y)=\frac{1}{\sqrt{2\pi R}}\varphi^{(0)}(x)+\sum_{n=1}^\infty \left[\varphi^{(n)+}(x)\cos\left( \frac{ny}{R}\right)+\varphi^{(n)-}(x)\sin\left(\frac{ny}{R}\right) \right]\, ,
\end{equation}
where the zero mode $\varphi^{(0)}(x)$ is identified as the corresponding four dimensional SM field and the $\varphi^{(n)\pm}$ ones are recognized as its KK excitations. Since, in general, not all the zero modes of the Fourier series have associated a SM counterpart, as it is the case of, for example, the fifth component of the gauge fields, it is desirable to eliminate some of these degrees of freedom by imposing extra symmetries acting on the fifth coordinate. One possibility is to demand that the fields of the theory obey some definite parity property under the reflection $y\to -y$. If we impose that the five--dimensional fields are even under reflection, only the zero mode and the $\varphi^{(n)+}$ coefficients appear in the corresponding Fourier series, whereas if we require that such fields are odd, only the $\varphi^{(n)-}$ components are present in the series. To implement this symmetry, one replaces the circle $S^1$ by the orbifold $S^1/Z_2$ in which $y$ is identified with $-y$. In this work, we will use this orbifold construction to eliminate some degrees of freedom.

As already mentioned, theories in more than four dimensions are not renormalizable in the Dyson's sense. Consequently, there is no limit for the number of extra--dimensional gauge invariants that can be introduced. In the context of the SM gauge group in five dimensions, $[{\rm SU_C}(3)\times {\rm SU_L}(2)\times {\rm U_Y}(1)]_5$, the effective action can be written as
\begin{equation}
S_{\rm eff}=\int d^4x\, {\cal L}^{\rm eff}_{\rm 4SM} \, ,
\end{equation}
with~\cite{NT,NT2}
\begin{equation}
{\cal L}^{\rm eff}_{\rm 4SM}=\int^{2\pi R}_0 dy\left[{\cal L}_{\rm 5SM}+\sum_N^\infty \frac{\beta_N g^{N_1}_5}{M^{N_2}_{\rm s}}{\cal O}_{N}\right]\, .
\end{equation}
 In the above expression, ${\cal L}_{\rm 5SM}$ is the five--dimensional version of the SM, whose structure in terms of KK modes will be discussed with some detail below. The ${\cal O}_N$ are operators of canonical dimension $N>5$, $M_{\rm s}$ is the energy scale above which the new physics first directly manifests itself, and $\beta_N$ is a dimensionless parameter that depends on the details of the underlying physics. In the above Lagrangian, it is assumed that all the independent operators that respect the five--dimensional Lorentz and gauge symmetries are included and that each of them is multiplied by an unknown dimensionless parameter $\beta_N$. The canonical dimension of each term of the series is appropriately corrected by introducing factors containing powers of the dimensionful coupling constants $g_5$ (the rule is to introduce a $g_5$ per each curvature appearing in the ${\cal O}_{N}$--invariant) and the $M_{\rm s}$ scale. Operators of higher canonical dimension will be more suppressed because they will involve higher powers of the cutoff scale $M_{\rm s}$.

We now proceed to discuss the diverse sectors of the five--dimensional generalization of the SM. The corresponding four--dimensional Lagragian can be written as follows:
 \begin{equation}
{\cal L}_{\rm 4SM}=\int^{2\pi R}_0 dy \, \left({\cal L}_{\rm 5YM}+{\cal L}_{\rm 5H}+{\cal L}_{\rm 5C}+{\cal L}_{\rm 5Y} \right)\, ,
\end{equation}
where ${\cal L}_{\rm 5YM}$, ${\cal L}_{\rm 5H}$, ${\cal L}_{\rm 5C}$, and ${\cal L}_{\rm 5Y}$ stand for the five--dimensional Yang--Mills, Higgs, Currents, and Yukawa sectors, respectively.

\subsection{The Yang--Mills sector}
We first discuss the Yang--Mills sector of the model. A comprehensive study of the gauge structure of the ${\rm SU}(N)$ compactified theory and its quantization was presented in~\cite{NT}, so in the present work we will include only those details that are essential for our discussion. This sector is given by
\begin{eqnarray}
\label{ymi}
{\cal L}_{\rm 4YM}&=&\int^{2\pi R}_0dy\, \left( -\frac{1}{4}{\cal G}^a_{MN}(x,y){\cal G}^{MN}_a(x,y)
-\frac{1}{4}{\cal W}^i_{MN}(x,y){\cal W}^{MN}_i(x,y)
\right.
\nonumber \\  &&
\left.
-\frac{1}{4}{\cal B}_{MN}(x,y){\cal B}^{MN}(x,y)\right)\, ,
\end{eqnarray}
where ${\cal G}^a_{MN}(x,y)$, ${\cal W}^i_{MN}(x,y)$, and ${\cal B}_{MN}(x,y)$ are the curvatures associated with the five--dimensional gauge groups ${\rm SU_C}(3)$, ${\rm SU_L}(2)$, and ${\rm U_Y}(1)$, respectively. As far as discrete indices are concerned, capital roman indices will run over the five space--time coordinates, whereas the usual four dimensional Lorentz indices will be denoted by $\mu,\nu,\ldots$. In addition, the symbols $a,b \cdots$ and $i,j\cdots $, will be used to denote gauge indices associated with the ${\rm SU_C}(3)$ and ${\rm SU_L}(2)$ groups, respectively. We assume that the ${\cal A}^a_\mu(x,y)$ (${\cal A}={\cal G}, {\cal W}, {\cal B}$) components of a five--dimensional gauge field ${\cal A}^a_{M}(x,y)$ are even when reflecting $y$ into $-y$, so they are KK--expanded as
\begin{equation}
{\cal A}^a_\mu(x,y)=\frac{1}{\sqrt{2\pi R}}A^{(0)a}_\mu(x)+\sum_{n=1}^\infty \frac{1}{\sqrt{\pi R}}A^{(n)a}_\mu(x)\cos\left(\frac{ny}{R}\right)\, .
\end{equation}
As to the fifth component ${\cal A}^a_5(x,y)$ of ${\cal A}^a_{M}(x,y)$, an odd parity under $y\rightarrow-y$ is assumed, so its Fourier series is
\begin{equation}
{\cal A}^a_5(x,y)=\sum_{n=1}^\infty \frac{1}{\sqrt{\pi R}}A^{(n)a}_5(x)\sin\left(\frac{ny}{R}\right)\,.
\end{equation}
This parity is required in order to avoid the presence of a zero mode of this component, whose existence would be associated with a physical scalar field without a SM counterpart. As emphasized in reference~\cite{NT}, the preservation of gauge invariance at the level of the four--dimensional theory relies on KK--expanding the curvatures instead of the gauge fields inside the integral sign in~(\ref{ymi}). This leads to the Lagrangian
\begin{eqnarray}
\label{lym}
{\cal L}_{\rm 4YM}&=&-\frac{1}{4}\left({\cal G}^{(0)a}_{\mu \nu}{\cal G}^{(0)a\mu \nu}+{\cal G}^{(n)a}_{\mu \nu}{\cal G}^{(n)a\mu \nu}+2{\cal G}^{(n)a}_{\mu 5}{\cal G}^{(n)a\mu 5}\right)\nonumber \\
&&-\frac{1}{4}\left({\cal W}^{(0)i}_{\mu \nu}{\cal W}^{(0)i\mu \nu}+{\cal W}^{(n)i}_{\mu \nu}{\cal W}^{(n)i\mu \nu}+2{\cal W}^{(n)i}_{\mu 5}{\cal W}^{(n)i\mu 5}\right)\nonumber \\
&&-\frac{1}{4}\left(B^{(0)}_{\mu \nu}B^{(0)\mu \nu}+B^{(n)}_{\mu \nu}B^{(n)\mu \nu}+2B^{(n)}_{\mu 5}B^{(n)\mu 5}\right)\, ,
\end{eqnarray}
where the KK modes indices are placed between parentheses. As usual, sums over repeated indices, including the modes ones, are assumed. The diverse covariant objects appearing in the above Lagrangian are given by
\begin{eqnarray}
{\cal F}^{(0)a}_{\mu \nu}&=&F^{(0)a}_{\mu \nu}+g f^{abc}A^{(n)b}_\mu A^{(n)c}_\nu \, ,  \\
{\cal F}^{(n)a}_{\mu \nu}&=&{\cal D}^{(0)ab}_\mu A^{(n)b}_\nu-{\cal D}^{(0)ab}_\nu A^{(n)b}_\mu +gf^{abc}\Delta^{nrs}A^{(r)b}_\mu A^{(s)c}_\nu \, ,  \\
{\cal F}^{(n)a}_{\mu 5}&=&{\cal D}^{(0)ab}_\mu A^{(n)b}_5 +\frac{n}{R}A^{(n)a}_\mu +gf^{abc}\Delta'^{nrs}A^{(r)b}_\mu A^{(s)c}_5 \, ,
\end{eqnarray}
for the non--Abelian gauge structures (${\cal F}^{(n)a}_{\mu \nu}={\cal G}^{(n)a}_{\mu \nu}$, ${\cal W}^{(n)i}_{\mu \nu}$), whereas for the Abelian group one has
\begin{eqnarray}
B^{(n)}_{\mu \nu}&=&\partial_\mu B^{(n)}_\nu-\partial_\nu B^{(n)}_\mu \, ,\\
B^{(n)}_{\mu 5}&=&\partial_\mu B^{(n)}_5+\frac{n}{R}B^{(n)}_\mu \, ,
\end{eqnarray}
where
\begin{eqnarray}
F^{(0)a}_{\mu \nu}&=&\partial_\mu A^{(0)a}_\nu-\partial_\nu A^{(0)a}_\mu+gf^{abc}A^{(0)b}_\mu A^{(0)c}_\nu \, , \\
B^{(0)}_{\mu \nu}&=&\partial_\mu B^{(0)}_\nu-\partial_\nu B^{(0)}_\mu \, ,
\end{eqnarray}
and
\begin{eqnarray}
\Delta^{nrs}&=&\frac{1}{\sqrt{2}}\left(\delta^{r,n+s}+\delta^{n,s+r}+\delta^{s,n+r} \right)\, ,\\
\Delta'^{nrs}&=&\frac{1}{\sqrt{2}}\left(\delta^{r,n+s}+\delta^{n,s+r}-\delta^{s,n+r} \right)\, .
\end{eqnarray}
In the above expressions, $g$ represents the dimensionless couplings associated with the non--Abelian  SM gauge groups and $A$ stands for $G$ or $W$. In addition, we have employed the following definitions:
\begin{eqnarray}
{\cal D}^{(0)ab}_\mu &=&\delta^{ab}\partial_\mu-gf^{abc}A^{(0)c}_\mu \, , \\
{\cal D}^{(nr)ab}_\mu &=&\delta^{nr}{\cal D}^{(0)ab}_\mu-gf^{abc}\Delta^{nsr}A^{(s)c}_\mu \, , \\
{\cal D}^{(nr)ab}_5&=&-\delta^{nr}\delta^{ab}\, \frac{n}{R}-gf^{abc}\Delta'^{nsr}A^{(s)c}_5\, .
\end{eqnarray}

In the context of UED, where the gauge parameters propagate in the fifth dimension, both the zero modes $A^{(0)a}_\mu$ and the KK ones $A^{(n)a}_\mu$ ($A=G,W,B$) are gauge fields, whereas the KK modes of the fifth component, $ A^{(n)a}_5$, are pseudo--Goldstone bosons~\cite{NT}. The zero modes of the gauge parameters define the well known infinitesimal standard gauge transformations (SGT), under which the non--Abelian fields $A^{(0)a}_\mu$ transform as gauge fields, whereas the KK excitations $A^{(n)a}_\mu$ and $A^{(n)a}_5$ transform under the adjoint representation of the group~\cite{NT}:
\begin{eqnarray}
\delta A^{(0)a}_\mu&=&{\cal D}^{(0)ab}_\mu \alpha^{(0)b}\, ,\\
\delta A^{(n)a}_\mu&=&gf^{abc}A^{(n)b}_\mu \alpha^{(0)c}\, ,\\
\delta A^{(n)a}_5&=&gf^{abc}A^{(n)b}_5 \alpha^{(0)c}\, .
\end{eqnarray}
The analogous infinitesimal SGT for the Abelian case are given by
\begin{eqnarray}
\delta B^{(0)}_\mu &=&\partial_\mu \alpha^{(0)}\, , \\
\delta B^{(n)}_\mu &=&0\, , \\
\delta B^{(n)}_5 &=&0\, .
\end{eqnarray}
On the other hand, the KK modes of the gauge parameters, $\alpha^{(n)a}$, define local non--standard gauge transformations (NSGT)~\cite{NT}. Under this sort of infinitesimal gauge transformations, the excited KK modes $A^{(n)a}_\mu$ transform as gauge fields~\cite{NT}, whereas the zero modes and pseudo--Goldstone bosons obey unusual laws of transformation:
\begin{eqnarray}
\delta A^{(0)a}_\mu &=&gf^{abc}A^{(n)b}_\mu \alpha^{(n)c}\, , \\
\delta A^{(n)a}_\mu &=&{\cal D}^{(nr)ab}_\mu \alpha^{(r)b}\, ,\\
\label{ta5}
\delta A^{(n)a}_5 &=&{\cal D}^{(nr)ab}_5 \alpha^{(r)b}\,.
\end{eqnarray}
The analogous NSGT for the Abelian case are
\begin{eqnarray}
\delta B^{(0)}_\mu &=&0\, , \\
\delta B^{(n)}_\mu &=&\partial_\mu \alpha^{(n)}\, , \\
\label{tb5}
\delta B^{(n)}_5 &=&-\frac{n}{R}\alpha^{(n)}\, .
\end{eqnarray}
Notice that, in the Abelian case, the laws of transformation of both $B^{(0)}_\mu$ and $B^{(n)}_\mu$ have the same form. Also, notice that the $B^{(n)}_5$ scalars transform trivially under the SGT but not under the NSGT, which is consistent with the fact that the latter type of gauge transformations is broken by compactification, which in turn implies that these fields are pseudo--Goldstone bosons. In fact, these scalar fields can be eliminated of the theory through a particular infinitesimal gauge transformation. Consider a NSGT with infinitesimal gauge parameters given by $\alpha^{(n)a}=(R/n)A^{(n)a}_5$ (for the non--Abelian gauge groups) and $\alpha^{(n)}=(R/n)B^{(n)}_5$ (for the Abelian group). Then, from Eqs. (\ref{ta5}) and (\ref{tb5}), we can see that $A^{(n)a}_5\to A'^{(n)a}_5=0$ and $B^{(n)}_5\to B'^{(n)}_5=0$. In this gauge, the terms that involve these scalar fields take the form
\begin{eqnarray}
\frac{1}{2}{\cal F}^{(n)a}_{\mu5}{\cal F}^{(n)a\mu}_5&=&\frac{1}{2}\left(\frac{n}{R}\right)^2 A^{(n)a}_\mu A^{(n)a\mu}\, , \\
\frac{1}{2}{\cal B}^{(n)}_{\mu 5}{\cal B}^{(n)\mu}_5&=&\frac{1}{2}\left(\frac{n}{R}\right)^2 B^{(n)}_\mu B^{(n)\mu}\, .
\end{eqnarray}
It is not difficult to prove that the ${\cal L}_{\rm 4YM}$ Lagrangian is separately invariant under both the SGT and the NSGT. The quantization of this theory was discussed in~\cite{NT}.

\subsection{The Higgs sector}
The Higgs sector is constituted by the kinetic term and the potential:
\begin{equation}
{\cal L}_{\rm H}=\int^{2\pi R}_0 dy\,\left[ (D_M\Phi)^\dag (x,y)(D^M \Phi)(x,y)- {\cal V}(\Phi^\dag,\Phi) \right]\, ,
\end{equation}
where an even parity is assumed for the five dimensional Higgs doublet, so that its corresponding KK expansion is:
\begin{equation}
\Phi(x,y)=\frac{1}{\sqrt{2\pi R}}\Phi^{(0)}(x)+\sum_{n=1}^\infty \frac{1}{\sqrt{\pi R}} \Phi^{(n)}(x)\cos\left(\frac{ny}{R}\right)\, .
\end{equation}
After expanding the covariant objects $(D_\mu \Phi)$ and $(D_5\Phi)$ in KK towers, and integrating out the fifth dimension, the kinetic term can be written as:
\begin{eqnarray}
{\cal L}_{\rm 4HK}&=&\int_0^{2\pi R}dy\hspace{0.1cm}(D_M\Phi)^\dag(x,y)(D^M\Phi)(x,y)
\nonumber \\ \nonumber \\&=&
(D_\mu\Phi)^{(0)\dag} (x)(D^\mu \Phi)^{(0)}(x)+(D_\mu\Phi)^{(n)\dag} (x)(D^\mu \Phi)^{(n)}(x)
\nonumber \\ &&
+(D_5\Phi)^{(n)\dag} (x)(D^5 \Phi)^{(n)}(x)\, ,\
\end{eqnarray}
where, as before, any pair of repeated indices, including the modes ones, indicate a sum. The four dimensional covariant objects $(D_\mu\Phi)^{(0)}$, $(D_\mu\Phi)^{(m)}$ and $(D_5\Phi)^{(m)}$, appearing in the above expression, are given by
\begin{eqnarray}
(D_\mu \Phi)^{(0)}&=&D^{(0)}_\mu \Phi^{(0)}-\left(ig\frac{\sigma^i}{2}W^{(n)i}_\mu+ig'\frac{Y}{2}B^{(n)}_\mu \right)\Phi^{(n)}\, ,\\
(D_\mu \Phi)^{(n)}&=&D^{(nr)}_\mu \Phi^{(r)}-\left(ig\frac{\sigma^i}{2}W^{(n)i}_\mu+ig'\frac{Y}{2}B^{(n)}_\mu \right)\Phi^{(0)}\, , \\
(D_5 \Phi)^{(n)}&=&D^{(nr)}_5\Phi^{(r)}-\left(ig\frac{\sigma^i}{2}W^{(n)i}_5+ig'\frac{Y}{2}B^{(n)}_5 \right)\Phi^{(0)}\, ,
\end{eqnarray}
where
\begin{eqnarray}
D^{(0)}_\mu &=&\partial_\mu-ig\frac{\sigma^i}{2}W^{(0)i}_\mu-ig'\frac{Y}{2}B^{(0)}_\mu \, , \\
D^{(nr)}_\mu&=&\delta^{nr}D^{(0)}_\mu-\Delta^{nsr}\left(ig\frac{\sigma^i}{2}W^{(s)i}_\mu+ig'\frac{Y}{2}B^{(s)}_\mu \right)\, , \\
D^{nr}_5&=&-\delta^{nr}\frac{n}{R}-\Delta^{'nsr}\left(ig\frac{\sigma^i}{2}W^{(s)i}_5+ig'\frac{Y}{2}B^{(s)}_5 \right)\, .
\end{eqnarray}
Under the electroweak group, $\Phi^{(0)}$ and $\Phi^{(n)}$ transform as:
\begin{eqnarray}
\delta \Phi^{(0)}&=&-\left(ig\frac{\sigma^i}{2}\alpha^{(0)i}+ig'\frac{Y}{2}\alpha^{(0)} \right)\Phi^{(0)}
-\left(ig\frac{\sigma^i}{2}\alpha^{(n)i}+ig'\frac{Y}{2}\alpha^{(n)} \right)\Phi^{(n)}\, ,
\label{GTKKzm}
\\
\delta \Phi^{(n)}&=&-\left(ig\frac{\sigma^i}{2}\alpha^{(0)i}+ig'\frac{Y}{2}\alpha^{(0)} \right)\Phi^{(n)}
-\left(ig\frac{\sigma^i}{2}\alpha^{(r)i}+ig'\frac{Y}{2}\alpha^{(r)} \right)\left(\delta^{nr}\Phi^{(0)}+\Delta^{nsr}\Phi^{(s)} \right)\, .
\label{GTKKex}
\end{eqnarray}
The infinitesimal SGT are obtained by taking $\alpha^{(n)i}=0=\alpha^{(n)}$ in Eqs.(\ref{GTKKzm}) and (\ref{GTKKex}), while the NSGT are derived when $\alpha^{(0)i}=0=\alpha^{(0)}$ in such expressions. It is assumed that $\Phi^{(0)}$ develops a vacuum expectation value (VEV), but the excited KK doublets $\Phi^{(n)}$ do not.

On the other hand, the Higgs potential is given by
\begin{equation}
V_4=\int^{2\pi R}_0 dy \left[\mu^2\left(\Phi^\dag(x,y) \Phi(x,y) \right)+\lambda_5\left(\Phi^\dag (x,y)\Phi(x,y) \right)^2\right] \, .
\end{equation}
Since the the Higgs doublet has canonical dimension $3/2$, $\mu$ and $\lambda_5$ have units of mass and inverse of mass, respectively. Once integrated out the fifth dimension, one obtains
\begin{eqnarray}
V_4&=&\mu^2\left(\Phi^{(0)\dag} \Phi^{(0)} \right)+\lambda\left(\Phi^{(0)\dag} \Phi^{(0)} \right)^2
+\left[\mu^2+2\lambda \left(\Phi^{(0)\dag}\Phi^{(0)}\right) \right]\left(\Phi^{(n)\dag}\Phi^{(n)} \right)\nonumber \\
&&+\lambda\left(\Phi^{(0)\dag}\Phi^{(n)}+\Phi^{(n)\dag}\Phi^{(0)} \right)\left(\Phi^{(0)\dag}\Phi^{(n)}+\Phi^{(n)\dag}\Phi^{(0)} \right)\nonumber \\
&&+2\lambda \Delta^{npq}\left(\Phi^{(0)\dag}\Phi^{(n)}+\Phi^{(n)\dag}\Phi^{(0)} \right)\left(\Phi^{(p)\dag}\Phi^{(q)} \right)
+\lambda \Delta^{npqr}
\left(\Phi^{(n)\dag}\Phi^{(p)} \right)\left(\Phi^{(q)\dag}\Phi^{(r)} \right)\, ,
\end{eqnarray}
with $\lambda=(\lambda_5/2\pi R)$ and
\begin{equation}
\Delta^{npqr}=\frac{1}{2}\left(\delta^{n,p+q+r}+\delta^{p,n+q+r}+\delta^{q,n+p+r}+\delta^{r,n+p+q}
+\delta^{n+p,q+r}
+\delta^{n+q,p+r}+\delta^{n+r,p+q} \right) \,.
\end{equation}
When the $\Phi^{(0)}$ Higgs doublet develops the vacuum expectation value $\Phi^\dag_0=(0,v/\sqrt{2})$, the KK zero modes of the theory acquire masses as it occurs in the SM, while the masses of the excited ones receive corrections at this scale. The details of this are presented in \ref{bm}.

\subsection{A covariant gauge--fixing procedure}
Now we turn to introduce a gauge--fixing procedure of renormalizable type. As it has been emphasized through the paper, the theory is invariant under two sets of infinitesimal gauge transformations~\cite{NT}: the SGT and the NSGT. Consequently, two gauge--fixing procedures must be introduced in order to define propagators for the zero KK gauge modes, $A^{(0)}_\mu$,  and the excited ones, $A^{(n)}_\mu$ ($A=G,W,B$). Due to the fact that the SGT are defined by the zero modes of the gauge parameters, whereas the NSGT depend exclusively on the excited ones, it is possible to introduce independent methods to remove the degeneration of the theory with respect to each of these sets of transformations~\cite{NT}. In the context of the SM, besides defining the gauge propagators, the $R_\xi$--gauges allow us to remove some bilinear terms of the Lagrangian that involve gauge and pseudo--Goldstone bosons, which arise as a consequence of a spontaneous symmetry breaking implemented in the theory. In our case, these types of terms arise from the Higgs mechanism and also from the compactification of the fifth dimension~\cite{NT}.

Since the excited KK Higgs doublets $\Phi^{(n)}$ do not develop a VEV, the Higgs mechanism is implemented in the standard way, via a VEV of the KK zero--mode doublet $\Phi^{(0)}$. In this case, the bilinear terms involving electroweak gauge bosons and their pseudo--Goldstone bosons are induced by the Higgs kinetic term
\begin{equation}
(D^{(0)}_\mu \Phi^{(0)})^\dag(D^{(0)\mu} \Phi^{(0)})=\left(D^{(0)}_\mu \Phi^{(0)}_0 \right)^\dag \left( D^{(0)\mu}\hat{\Phi}^{(0)}\right)+{\rm H.\, c.}+\cdots \, ,
\end{equation}
where $\Phi^{(0)\dag }_0=(0,v/\sqrt{2})$ and $\hat{\Phi}^{(0)}=\Phi^{(0)}-\Phi^{(0)\dag }_0$. Diverse gauge--fixing procedures that remove the degeneration with respect to the SGT transformations and eliminate this bilinear term are well known in the literature. Here, we limit our discussion to a gauge--fixing procedure that is covariant under the electromagnetic gauge group and considerably simplifies the calculations of loop amplitudes~\cite{HT}. Under such circumstances, the fixation of the gauge for the SGT is determined by the following gauge--fixing functions:
\begin{itemize}

\item ${\rm  SU_C}(3)$
\begin{equation}
f^{(0)a}=\partial_\mu G^{(0)a\mu} \, ,
\end{equation}

\item ${\rm SU_L}(2)$
\begin{eqnarray}
f^{(0)i}&=&\left(\delta^{ij}\partial_\mu -g'\epsilon^{ij3}B^{(0)}_\mu\right)W^{(0)j\mu}\nonumber +\frac{ig\xi}{2}\bigg[\Phi^{(0)\dag}\left(\sigma^i-i\epsilon^{ij3}\sigma^j\right)\Phi^{(0)}_0
\nonumber \\&&
-\Phi^{(0)\dag}_0\left(\sigma^i+i\epsilon^{ij3}\sigma^j\right)\Phi^{(0)}+i\epsilon^{ij3}\Phi^{(0)\dag}\sigma^j \Phi^{(0)}\bigg] \, ,
\end{eqnarray}

\item ${\rm U_Y}(1)$
\begin{equation}
f^{(0)}=\partial_\mu B^{(0)\mu}+\frac{ig\xi}{2}\left(\Phi^{(0)\dag}\Phi^{(0)}_0-\Phi^{(0)}_0\Phi^{(0)}\right)\, .
\end{equation}

\end{itemize}
It is not difficult to convince ourselves that this gauge--fixing procedure for the electroweak sector allows us to define the $W$ propagator in a covariant way under the ${\rm U}_e(1)$ group.

On the other hand, the compactification of the fifth dimension produces bilinear terms that involve gauge fields $A^{(n)}_\mu$ and scalar fields $A^{(n)}_5$ ($A=G,W,B)$. These terms arise from
\begin{equation}
\label{b1}
-\frac{1}{2}\left({\cal G}^{(n)a}_{\mu 5}{\cal G}^{(n)a\mu 5}+{\cal W}^{(n)i}_{\mu 5}{\cal W}^{(n)i\mu 5}+{\cal B}^{(n)}_{\mu 5}{\cal B}^{(n)\mu 5} \right)\, .
\end{equation}
In addition, the Higgs mechanism induces bilinear terms among the charged component $\phi^{(n)\pm}$ and the imaginary component $\phi^{(n)}_I$ of the scalar  doublet $\Phi^{(n)}$ with the scalar fields $W^{(n)\pm}_5$ and $W^{(n)3}_5$, respectively. These terms arise specifically from
\begin{equation}
\label{b2}
(D_5\Phi)^{(n)\dag}(D^5\Phi)^{(n)}\, .
\end{equation}
In Ref. \cite{NT}, we showed how to eliminate the bilinear terms appearing in Eq.(\ref{b1}) through a gauge--fixing procedure for the NSGT that is covariant under the SGT. Here, we generalize this method to cancel the bilinear terms of expression (\ref{b2}) as well. To this end, we introduce the following gauge--fixing procedure for the NSGT:
\begin{itemize}

\item ${\rm SU_C}(3)$
\begin{equation}
f^{(n)a}={\cal D}^{(0)ab}_\mu G^{(n)b\mu}-\xi \left(\frac{n}{R} \right)G^{(n)a}_5\, .
\end{equation}

\item ${\rm SU_L}(2)$
\begin{equation}
f^{(n)i}={\cal D}^{(0)ij}_\mu W^{(n)j\mu}-\xi \left(\frac{n}{R} \right)W^{(n)i}_5
+ig\xi\left(\Phi^{(n)\dag}\frac{\sigma^i}{2}\Phi^{(0)}-\Phi^{(0)\dag}\frac{\sigma^i}{2}\Phi^{(n)} \right) \, .
\end{equation}

\item ${\rm U_Y}(1)$
\begin{equation}
f^{(n)}=\partial_\mu B^{(n)\mu}-\xi \left(\frac{n}{R} \right)B^{(n)}_5 \,
+ig'\xi\left(\Phi^{(n)\dag}\frac{Y}{2}\Phi^{(0)}-\Phi^{(0)\dag}\frac{Y}{2}\Phi^{(n)} \right) \, .
\end{equation}

\end{itemize}
This gauge--fixing procedure for the the NSGT is covariant under the SGT. In fact, the gauge--fixing functions $f^{(n)a}$ and $f^{(n)i}$ transform under the adjoint representation of ${\rm SU_C}(3)$ and ${\rm SU_L}(2)$, respectively, whereas $f^{(n)}$ is invariant under ${\rm U_Y}(1)$.

The quantum theory of gauge systems is governed by the BRST symmetry~\cite{BRST}, which, at the classical level, emerges naturally in the context of the field--antifield formalism~\cite{AFAB}. The extended action in five dimensions can be written as follows~\cite{NT}:
\begin{eqnarray}
S&=&\int d^4x \int dy \Big[{\cal L}_{\rm 5SM}+\sum_{{\cal A}={\cal G,W}}\left({\cal A}^{*}_{Ma}{\cal D}^{abM} {\cal C}^b+\frac{1}{2}g_5f^{abc}{\cal C}^*_c{\cal C}^b{\cal C}^a+\bar{{\cal C}}^{*a}B_a \right)\nonumber \\
&&+\Phi^*\frac{\sigma^i}{2} \delta \Phi +{\cal B}^*_M\partial^M {\cal C} +\bar{{\cal C}}^*B\Big] \, ,
\end{eqnarray}
where the asterisk denotes antifields, whereas the ${\cal C}$ fields stand for the Faddeev--Popov ghosts. Additionally, the $\bar{{\cal C}}$ and $B$ fields are known as trivial pairs, with the former representing the well known Faddeev--Popov antighosts and the latter incarnating the so--called auxiliary fields. As we will see later, the presence of such objects is needed to remove the degeneration associated to gauge symmetry. In the above expression, the last two terms correspond to the Abelian group. It is important to note that we have included the antifields for the Higgs doublet, which are needed in theories with spontaneous symmetry breaking in order to introduce a gauge--fixing procedure that allows us to generate unphysical masses for pseudo-Goldstone bosons and ghosts. Within this formalism, the compactified extended action~\cite{NT}, which is a functional of fields and antifields, is given by
\begin{eqnarray}
S&=&\int d^4x\Big\{ {\cal L}_{\rm 4SM}+\sum_{A=G,W}\Big[A^{(0)*}_{\mu a}{\cal D}^{(0)ab\mu} C^{(0)b}
+\frac{1}{2}gf^{abc}C^{(0)*}_c C^{(0)b}C^{(0)a}
+\bar{C}^{(0)*a}B^{(0)}_a
\nonumber \\ &&
+ A^{(m)*}_{\mu a}{\cal D}^{(mn)ab\mu}C^{(n)b}
-A^{(m)*}_{5a}{\cal D}^{(mn)ab}C^{(n)b}
+\frac{1}{2}gf^{abc}C^{(0)*}_cC^{(m)b}C^{(m)a}
\nonumber \\&&
+\frac{1}{2} f^{abc}C^{(m)*}_c\left( C^{(0)b}C^{(m)a}+C^{(0)a}C^{(m)b}+\Delta^{mrn}C^{(r)b}C^{(n)a}\right)
+\bar{C}^{(m)*a}B^{(m)}_a\Big]
\nonumber \\ &&
+\Phi^{(0)*}\delta \Phi^{(0)}+\Phi^{(m)*}\delta \Phi^{(m)}
+B^{(0)*}_\mu \partial^\mu C^{(0)}
+\bar{C}^{(0)*}B^{(0)}+B^{(m)*}_\mu \partial^\mu C^{(m)}
\nonumber \\ &&
-\frac{m}{R}B^{(m)}_5 C^{(m)}+\bar{C}^{(m)*}B^{(m)}
\Big \}\, ,
\end{eqnarray}
where
\begin{equation}
\delta \Phi^{(0)}=ig\frac{\sigma^i}{2}\Phi^{(0)}C^{(0)i}+g'\frac{Y}{2}\Phi^{(0)}C^{(0)} \, ,
\end{equation}
\begin{eqnarray}
\delta \Phi^{(m)}&=&ig\frac{\sigma^i}{2}\left(\Phi^{(0)}C^{(m)i}+\Phi^{(m)}C^{(0)i}+\Delta^{mrs}\Phi^{(r)}C^{(s)i}\right)
\nonumber \\ &&
+ig'\frac{Y}{2}\left(\Phi^{(0)}C^{(m)}+\Phi^{(m)}C^{(0)}+\Delta^{mrs}\Phi^{(r)}C^{(s)}\right)\, .
\end{eqnarray}
The above action still possesses gauge invariance, so that a gauge--fixing procedure must be introduced to lift such degeneration. On the other hand, the antifields do not represent true degrees of freedom, for which they must be removed of the theory before quantizing it. The gauge--fixing procedure is determined through a criterion that permits us to remove the antifields of the theory in a nontrivial way (they cannot be just set to zero). The criterion that allows one to eliminate the antifields and, at the same time, to lift the degeneration consists in the following: one introduces a fermionic functional of the fields, $\Psi[\Phi]$, with ghost number -1 and such that, for a given antifield $\Phi^*_A$, it satisfies
\begin{equation}
\label{fermionic}
\Phi^*_A=\frac{\partial \Psi}{\partial \Phi^A} \, .
\end{equation}
 Notice that the presence of the trivial pairs, $\bar{C}^a$  and $B_a$, is necessary since the only fields with ghost number $-1$ are precisely the antighosts. To lift the degeneration with respect to the SGT, we introduce the following fermionic functional
\begin{eqnarray}
\Psi_{\rm SGT}&=&\int d^4x \Big[\bar{C}^{(0)a}\left(f^{(0)a}+\frac{\xi}{2}B^{(0)a}+gf^{abc}\bar{C}^{(0)b}C^{(0)c} \right) \nonumber \\
&&+\bar{C}^{(0)i}\left(f^{(0)i}+\frac{\xi}{2}B^{(0)i}+g\epsilon^{ijk}\bar{C}^{(0)j}C^{(0)k} \right)+\bar{C}^{(0)}\left(f^{(0)}+\frac{\xi}{2}B^{(0)} \right)\Big] \, .
\end{eqnarray}
On the other hand, we can to lift the degeneration associated with the NSGT through the fermionic functional
\begin{eqnarray}
\Psi_{\rm NSGT}&=&\int d^4x \Big[\bar{C}^{(m)a}\left(f^{(m)a}+\frac{\xi}{2}B^{(m)a}+gf^{abc}\Delta^{mrn}\bar{C}^{(r)b}C^{(n)c} \right) \nonumber \\
&&+\bar{C}^{(m)i}\left(f^{(m)i}+\frac{\xi}{2}B^{(m)i}+g\epsilon^{ijk}\Delta^{mrn}\bar{C}^{(r)j}C^{(n)k} \right)+\bar{C}^{(m)}\left(f^{(m)}+\frac{\xi}{2}B^{(m)} \right)\Big] \, .
\end{eqnarray}
Once removed the antifields, via Eq.(\ref{fermionic}), and eliminated the auxiliary fields, by using the equations of motion, one obtains the gauge--fixed BRST action
\begin{equation}
S_\Psi=\int d^4x\Big[{\cal L}_{\rm 4SM}+{\cal L}_{\rm GF}+{\cal L}_{\rm FPG} \Big]\, ,
\end{equation}
where ${\cal L}_{\rm GF}$ is the gauge--fixing term, given by
\begin{equation}
{\cal L}_{\rm GF}=-\frac{1}{2\xi}\sum_{A=G,W}\left(f^{(0)a}f^{(0)a}+f^{(m)a}f^{(m)a}\right)-\frac{1}{2\xi}\left(f^{(0)}f^{(0)}+f^{(m)}f^{(m)}\right)\, ,
\end{equation}
and ${\cal L}_{\rm FPG}$ is the Faddeev--Popov ghost Lagrangian, which can be decomposed into two parts as
\begin{equation}
{\cal L}_{\rm FPG}={\cal L}^1_{\rm FPG}+{\cal L}_{\rm FPG}^2\, ,
\end{equation}
where ${\cal L}^1_{\rm FPG}$ contains the interactions between gauge and ghost fields:
\begin{eqnarray}
{\cal L}^1_{\rm FPG}&=& \sum_{A=G,W}\Big[\bar{C}^{(0)c}\frac{\partial f^{(0)c}}{\partial A^{(0)a}_\mu} {\cal D}^{(0)ab\mu}C^{(0)b}-\frac{2g}{\xi}f^{abc}f^{(0)a}\bar{C}^{(0)b}C^{(0)c}\nonumber \\
&&+\bar{C}^{(m)c}\left(\frac{\partial f^{(m)c}}{\partial A^{(n)a}_\mu}{\cal D}^{(nr)ab\mu}-\frac{\partial f^{(m)c}}{\partial A^{(n)a}_5}{\cal D}^{(nr)ab}_5 \right)C^{(r)b}-\frac{2g}{\xi}f^{abc}\Delta^{mrn}f^{(m)a}\bar{C}^{(r)b}C^{(n)c}\Big]
\nonumber \\ &&
+\left(\bar{C}^{(0)j}\frac{\partial f^{(0)j}}{\partial \Phi^{(0)}_i}+\bar{C}^{(0)}\frac{\partial f^{(0)}}{\partial \Phi^{(0)}_i}\right)\delta \Phi^{(0)}_i +\left(\bar{C}^{(n)j}\frac{\partial f^{(n)j}}{\partial \Phi^{(m)}_i}+\bar{C}^{(n)}\frac{\partial f^{(n)}}{\partial \Phi^{(m)}_i}\right)\delta \Phi^{(m)}_i \nonumber \\&&
+\bar{C}^{(0)}\frac{\partial f^{(0)}}{\partial B^{(0)}_\mu}\partial^\mu C^{(0)}+\bar{C}^{(m)}\frac{\partial f^{(m)}}{\partial B^{(n)}_\mu}\partial^\mu C^{(n)}\, .
\end{eqnarray}
On the other hand, ${\cal L}^2_{\rm FPG}$ contains only quartic interactions among ghost fields,
\begin{eqnarray}
{\cal L}^2_{\rm FPG}&=&\frac{g^2}{2}f^{abc}f^{cde}\sum_{A=G,W}\Big\{\bar{C}^{(0)d}\bar{C}^{(0)e}\Big[\left(1-\frac{4}{\xi}\right)C^{(0)b}C^{(0)a}
+C^{(m)a}C^{(m)b}\Big]
\nonumber \\&&
+\Delta^{mpq}\bar{C}^{(p)d}\bar{C}^{(q)e}\left(C^{(0)b}C^{(m)a}
\right.
\left.
+C^{(0)a}C^{(m)b}+\Delta^{mrn}C^{(r)b}C^{(n)a} \right)  \Big\}\, .
\end{eqnarray}

\subsection{The Currents and Yukawa sectors}
In five dimensions, the Dirac fields are still objects with four components, as in the four--dimensional case. This is due to the fact that the standard Dirac matrices $\Gamma^M=\gamma^\mu,\, i\gamma_5$ satisfy the Clifford algebra
\begin{equation}
\left[\Gamma^M,\, \Gamma^N\right]_+=2g^{MN}\, ,
\end{equation}
where $[,]_+$ stands for the anticommutator and $g^{MN}=(+\, -\, -\, -\, -)$ is the five--dimensional metric tensor. However, there is no chirality in five dimensions. The reason is that it is impossible to construct a nilpotent $\Gamma^5$ matrix that, in addition, anticommute with all the $\Gamma^M$. So, in five dimensions, no interactions of the gauge bosons associated with the ${\rm SU}(2)$ gauge group distinguishing chirality can be constructed. Fortunately, the $y\to -y$  parity operation can be used to reproduce the left--handed doublets and right--handed singlets of ${\rm SU_L}(2)$ at the four--dimensional level. Under this symmetry operation, the five--dimensional Dirac fields transform as $\psi \to \gamma_5 \psi(x,-y)$. Taking into account this and the fact that in four dimensions  right--handed fermions appear only as ${\rm SU_L}(2)$--singlets, whereas left--handed fermions are present only as ${\rm SU_L}(2)$--doublets, we demand that the corresponding five--dimensional representations of this group, $f(x,y)$ and $F(x,y)$, are, respectively, even and odd under this transformation. Accordingly, one can write
\begin{eqnarray}
f(x,y)&=&\frac{1}{\sqrt{2\pi R}}f^{(0)}_R(x)
+\sum_{n=1}^{\infty}\frac{1}{\sqrt{\pi R}}\left[ \hat{f}^{(n)}_R(x)\cos\left(\frac{ny}{R}\right)+\hat{f}^{(n)}_L(x)\sin\left(\frac{ny}{R}\right)\right] \, , \\ \nonumber \\
F(x,y)&=&\frac{1}{\sqrt{2\pi R}}F^{(0)}_L(x)
+\sum_{n=1}^{\infty}\frac{1}{\sqrt{\pi R}}\left[ F^{(n)}_L(x)\cos\left(\frac{ny}{R}\right)+F^{(n)}_R(x)\sin\left(\frac{ny}{R}\right)\right] \, .
\end{eqnarray}
The zero mode $f^{(0)}_R(x)$ represents one of the SM right--handed ${\rm SU_L}(2)$--singlets: a charged lepton $e^{(0)}(x)$, a quark of type up $u^{(0)}(x)$ or a quark of type down $d^{(0)}(x)$. In addition, the zero mode $F^{(0)}_L(x)$ stands for a SM left--handed lepton doublet $L^{(0)}(x)$ or a quark doublet $Q^{(0)}(x)$. The KK modes  $\hat{f}^{(n)}_L\left(\hat{f}^{(n)}_R\right)$ stand for left--handed (right--handed) ${\rm SU_L}(2)$--singlets, whereas $F^{(n)}_L\left(F^{(n)}_R\right)$ represents left--handed (right--handed) ${\rm SU_L}(2)$--doublets. Let us establish our notation as follows:
\begin{equation}
F^{(0)}_L=\left(\begin{array}{ccc}
 f^{(0)}_{uL}  \\
\, \, \\
f^{(0)}_{dL}
\end{array}\right) \, , \, \, F^{(n)}_L=\left(\begin{array}{ccc}
 f^{(n)}_{uL}  \\
\, \, \\
f^{(n)}_{dL}
\end{array}\right) \, , \, \, F^{(n)}_R=\left(\begin{array}{ccc}
 f^{(n)}_{uR}  \\
\, \, \\
f^{(n)}_{dR}
\end{array}\right) \, ,
\end{equation}
where the subscript $u$ denotes a neutrino or quark of type up, while $d$ stands for charged leptons or quarks of type down. As we will see below, after compactification, the SM fermions are given by $f^{(0)}=f^{(0)}_L+f^{(0)}_R$. On the other hand, the excited KK modes define massive states of type left, $F^{(n)}_L+F^{(n)}_R$, and right, $\hat{f}^{(n)}_L+\hat{f}^{(n)}_R$. It is an interesting feature of theories formulated in compactified extra dimensions that the mass terms corresponding to the excited KK modes are invariant under the four--dimensional gauge group.

The Currents sector is given by the Lagrangian
\begin{eqnarray}
{ \cal L}_{\rm 4C}&=&\int^{2\pi R}_0\Big[\sum_{L_1,L_2,L_3}i\bar{L}(x,y)\Gamma^M D_M L(x,y)
+\sum_{e,\mu,\tau}i\bar{e}(x,y)\Gamma^M D_M e(x,y)
\nonumber \\&&
+\sum_{Q_1,Q_2,Q_3}i\bar{Q}(x,y)\Gamma^M D_M Q(x,y)
+\sum_{u,c,t}i\bar{u}(x,y)\Gamma^M D_M u(x,y)
\nonumber \\ &&
+\sum_{d,s,b}i\bar{d}(x,y)\Gamma^M D_M d(x,y)  \Big] \, ,
\end{eqnarray}
where
\begin{equation}
D_M=\partial_M-ig_{\rm s5}\frac{\lambda^a}{2}{\cal G}^a_M-ig_5\frac{\sigma^i}{2}{\cal W}^{i}_M-ig'_5 \frac{Y}{2}{\cal B}_M \, ,
\end{equation}
with $\lambda^a$ representing the Gell--Man matrices. After expanding in Fourier series and integrating out the $y$ coordinate, the Currents sector can be written as follows:
\begin{eqnarray}
\label{currents}
{ \cal L}_{\rm 4C}=&\displaystyle\sum_{L_1,L_2,L_3,
Q_1,Q_2,Q_3}\Big[&i\bar{F}^{(0)}_L\gamma^\mu (D_\mu F)^{(0)}_L +i\bar{F}^{(n)}_L\gamma^\mu (D_\mu F)^{(n)}_L
+i\bar{F}^{(n)}_R\gamma^\mu (D_\mu F)^{(n)}_R
 \nonumber \\
&&
-\bar{F}^{(0)}_L(D_5F)^{(0)}_L
-\bar{F}^{(n)}_L (D_5 F)^{(n)}_L+\bar{F}^{(n)}_R (D_5 F)^{(n)}_R\Big]\nonumber \\ \nonumber \\
&\displaystyle+\sum_{e,\mu,\tau,d,s,b,u,c,t}\Big[&i\bar{f}^{(0)}_R\gamma^\mu (D_\mu \hat{f})^{(0)}_R +i\bar{\hat{f}}\hspace{0.001cm}^{(n)}_R\gamma^\mu (D_\mu \hat{f})^{(n)}_R
+i\bar{\hat{f}}\hspace{0.001cm}^{(n)}_L\gamma^\mu (D_\mu \hat{f})^{(n)}_L
\nonumber \\&&
-\bar{f}^{(0)}_R(D_5\hat{f})^{(0)}_R
-\bar{\hat{f}}\hspace{0.001cm}^{(n)}_R (D_5 \hat{f})^{(n)}_R+\bar{\hat{f}}\hspace{0.001cm}^{(n)}_L (D_5 \hat{f})^{(n)}_L\Big]\, .
\end{eqnarray}
The diverse covariant objects appearing in this expression are listed in \ref{A}.

The five--dimensional Yukawa sector is given by
\begin{eqnarray}
-{\cal L}_{\rm 4Y}=\int^{2\pi R}_0dy\Big[&&+ \sum_{\rm  families} \lambda_{5e} \, \bar{L}(x,y)e(x,y)\Phi(x,y)+\, {\rm H. \, c.}
\nonumber \\ &&
+\sum_{\rm  families} \lambda_{5d} \, \bar{Q}(x,y)d(x,y)\Phi(x,y)+\, {\rm H. \, c.} \nonumber \\
&&+\sum_{\rm  families} \lambda_{5u} \, \bar{Q}(x,y)u(x,y)\tilde{\Phi}(x,y)+\, {\rm H. \, c.} \Big]\, ,
\end{eqnarray}
where $\tilde{\Phi}(x,y)=i\sigma^2 \Phi^*(x,y)$. Integrating out the $y$ coordinate, one obtains
\begin{eqnarray}
-{\cal L}_{\rm 4Y}=&\displaystyle\sum_{\rm families}\lambda_{e}\Big[\bar{L}^{(0)}_Le^{(0)}_R\Phi^{(0)}+\left(\bar{L}^{(n)}_L\hat{e}^{(n)}_R+ \bar{L}^{(n)}_R\hat{e}^{(n)}_L\right)\Phi^{(0)}
+\left(\bar{L}^{(0)}_L\hat{e}^{(n)}_R+\bar{L}^{(n)}_Le^{(0)}_R \right)\Phi^{(n)}\nonumber \\ \nonumber \\
&+\left(\Delta^{nrs}\bar{L}^{(n)}_L\hat{e}^{(r)}_R
+\Delta'^{snr}\bar{L}^{(n)}_R\hat{e}^{(r)}_L \right)\Phi^{(s)}\Big]+ \, {\rm H. \, c.}\nonumber \\ \nonumber \\
&\displaystyle+\sum_{\rm  families}\lambda_{d}\Big[\bar{Q}^{(0)}_Ld^{(0)}_R\Phi^{(0)}+\left(\bar{Q}^{(n)}_L\hat{d}^{(n)}_R+ \bar{Q}^{(n)}_R\hat{d}^{(n)}_L\right)\Phi^{(0)}
+\left(\bar{Q}^{(0)}_L\hat{d}^{(n)}_R+\bar{Q}^{(n)}_Ld^{(0)}_R \right)\Phi^{(n)}\nonumber \\ \nonumber \\
&+\left(\Delta^{nrs}\bar{Q}^{(n)}_L\hat{d}^{(r)}_R
+\Delta'^{snr}\bar{Q}^{(n)}_R\hat{d}^{(r)}_L \right)\Phi^{(s)} \Big]+ \, {\rm H. \, c.}\nonumber \\ \nonumber \\ &
\displaystyle+\sum_{\rm  families}\lambda_{u}\Big[\bar{Q}^{(0)}_Lu^{(0)}_R\tilde{\Phi}^{(0)}+\left(\bar{Q}^{(n)}_L\hat{u}^{(n)}_R+ \bar{Q}^{(n)}_R\hat{u}^{(n)}_L\right)\tilde{\Phi}^{(0)}
+\left(\bar{Q}^{(0)}_L\hat{u}^{(n)}_R+\bar{Q}^{(n)}_Lu^{(0)}_R \right)\tilde{\Phi}^{(n)}\nonumber \\ \nonumber \\
&+\left(\Delta^{nrs}\bar{Q}^{(n)}_L\hat{u}^{(r)}_R
+\Delta'^{snr}\bar{Q}^{(n)}_R\hat{u}^{(r)}_L \right)\tilde{\Phi}^{(s)}\Big] + \, {\rm H. \, c.}
\end{eqnarray}
where $\lambda_i=\lambda_{5i}/\sqrt{2\pi R}$.

The Currents sector generates masses proportional to $n/R$ for the excited fermion KK modes, whereas in the Yukawa sector the Higgs mechanism endows the zero modes with mass at the same time that induces corrections to the gauge masses of the excited KK states. The term of the Yukawa sector involving only zero modes is diagonalized in the standard way, \textit{i.e.}, the flavor space vectors $N^{(0)}_L$ (neutrinos), $E^{(0)}_{L,R}$ (charged leptons), $U^{(0)}_{L,R}$ (quarks of up type), and $D^{(0)}_{L,R}$ (quarks of down type) are transformed into mass eigenstates via the unitary matrices $V^e_L$, $V^e_{L,R}$, $V^u_{L,R}$, and $V^d_{L,R}$, respectively. As far as the KK excitations are concerned, we demand that the corresponding KK vectors $N^{(n)}_{L,R}$, $E^{(n)}_{L,R}$, $U^{(n)}_{L,R}$, and $D^{(n)}_{L,R}$ transform through $V^e_L$, $V^e_L$, $V^u_L$, and $V^d_L$, respectively. On the other hand, in the case of the $\hat{E}^{(n)}_{L,R}$, $\hat{U}^{(n)}_{L,R}$, and $\hat{D}^{(n)}_{L,R}$ partners, we impose the condition that they transform via the matrices $V^e_R$, $V^u_R$, and $V^d_R$. With this choice no new sources of family flavor changing are present. A complete analysis of this is presented in \ref{fm}.

\section{The vertices of theory}
\label{vt}
One important feature of UED models is the conservation of the momentum in the extra dimensions. As a consequence, at tree level one can only couple two or more  KK excitations to a zero--mode field (SM field). This means that electroweak observables are not sensitive to virtual effects of KK excitations at tree level, but only at the one--loop level or higher orders. This is the reason why experimental constraints on UED models are quite weak~\cite{ACD}. Although the one--loop effects of KK excitations on SM Green's functions (light Green's functions) are determined by trilinear and quartic couplings of the way $\varphi^{(0)}\varphi^{(n)}\varphi^{(n)}$ and $\varphi^{(0)} \varphi^{(0)}\varphi^{(n)}\varphi^{(n)}$, the purpose of this section is to present a complete list of all the physical vertices of the theory.

\subsection{Scalar selfcouplings}
Couplings among scalars only can arise from the Higgs potential. The only SM Higgs field is the Higgs boson $H^{(0)}$. The rest of the physical scalars (see  \ref{bm}) are just KK excitations, namely, the charged scalars $H^{(n)\pm}$, the KK excitations of $H^{(0)}$, denoted by $H^{(n)}$, and the neutral CP--odd scalars  $A^{(n)}$. The scalar vertices that can contribute to SM Green's functions at the one--loop level are given by the Lagrangian
\begin{eqnarray}
{\cal L}^1_{\rm S}&=&-\frac{g m^2_{H^{(0)}}}{4m_{W^{(0)}}}H^{(0)}\left(H^{(0)}H^{(0)}+ \frac{3}{2}H^{(n)}H^{(n)}
+2c^2_\alpha H^{(n)-}H^{(n)+}+c^2_\beta A^{(n)}A^{(n)}\right)\nonumber \\
&&-\frac{g^2m^2_{H^{(0)}}}{32m^2_{W^{(0)}}}H^{(0)}H^{(0)}\left(H^{(0)}H^{(0)} +3H^{(n)}H^{(n)}
+4c^2_\alpha H^{(n)-}H^{(n)+}+2c^2_\beta A^{(n)}A^{(n)}\right)\, ,
\end{eqnarray}
where the SM couplings were included. Although the rest of the couplings among scalars do not contribute to the SM Green's functions at the one--loop level, they can do it at higher orders or impact nonstandard observables through
\begin{equation}
{\cal L}^2_{\rm S}=-\left(\frac{gm^2_{H^{(0)}}}{2m_{W^{(0)}}}+\frac{g^2m^2_{H^{(0)}}}{4m^2_{W^{(0)}}}H^{(0)}\right)\Delta^{nrs}H^{(n)}\left[c^2_\alpha H^{(r)-}H^{(s)+}
+\frac{1}{2}\left(H^{(r)}H^{(s)}+c^2_\beta A^{(r)}A^{(s)}\right)\right]\, ,
\end{equation}

\begin{eqnarray}
{\cal L}^3_{\rm S}&=&-\frac{g^2m^2_{H^{(0)}}}{8m^2_{W^{(0)}}}\Delta^{npqr}\left[c^2_\alpha H^{(n)-}H^{(p)+}+\frac{1}{2}\left(H^{(n)}H^{(p)}+c^2_\beta A^{(n)}A^{(p)}\right)\right] \nonumber \\
&&\times\left[c^2_\alpha H^{(q)-}H^{(r)+}+\frac{1}{2}\left(H^{(q)}H^{(r)}+c^2_\beta A^{(q)}A^{(r)}\right)\right] \, .
\end{eqnarray}
It is important to keep in mind our convention of sum over repeated KK indices, which run from 1 to infinity. This notation will be systematically used through the paper.

\subsection{Scalar--gauge boson couplings}
The couplings among physical scalars ($H^{(0)}, H^{(n)\pm}, H^{(n)}, A^{(n)}$) and gauge bosons arise from the kinetic term of the Higgs sector and also from the Yang--Mills term $\frac{1}{2}{\cal W}^{(n)-}_{\mu 5}{\cal W}^{(n)+\mu}_{\ \ \ \ \ \ \ 5}$. The last is due to the fact that $\phi^{(n)\pm}$ and $W^{(n)\pm}_5$ mix to produce the mass eigenstate $H^{(n)\pm}$ and the pseudo--Goldstone boson $G^{\pm}_{W^{(n)}}$ (see \ref{bm}). Similarly, the $\phi^{(n)}_I-W^{(n)3}_5$ mixing leads to the physical CP--odd mass--eigenstate scalar $A^{(n)}$ and to the pseudo--Goldstone boson $G_{Z^{(n)}}$. The Lagrangian of this sector can conveniently be reorganized as follows:
\begin{eqnarray}
\label{KT}
{\cal L}_{Kinetic}&=&(D_\mu \Phi)^{(n)\dag}(D^\mu \Phi)^{(n)}+(D_5 \Phi)^{(n)\dag}(D^5 \Phi)^{(n)}\nonumber \\
&&+{\cal W}^{(n)+}_{\mu 5}{\cal W}^{(n)-\mu}_{\ \ \ \ \ \ \ 5}+\frac{1}{2}{\cal W}^{(n)3}_{\mu 5}{\cal W}^{(n)3 \mu}_{\ \ \ \ \ \ 5}+\frac{1}{2}B^{(n)}_{\mu 5}B^{(n)\mu}_{\ \ \ \ \ 5}\, ,
\end{eqnarray}
where
\begin{eqnarray}
(D_\mu \Phi)^{(n)}&=&\partial_\mu \Phi^{(n)}-ig\left({\cal O}^{(0)}_\mu \Phi^{(n)}+ {\cal O}^{(n)}_\mu \Phi^{(0)}+\Delta^{nrs}{\cal O}^{(s)}_\mu \Phi^{(r)}\right) \, , \\
(D_5 \Phi)^{(n)}&=&-\frac{n}{R}\Phi^{(n)}-ig\left({\cal O}^{(n)}_5 \Phi^{(0)}+\Delta'^{nrs}{\cal O}^{(s)}_5 \Phi^{(r)}\right)\, ,
\end{eqnarray}
with
\begin{eqnarray}
{\cal O}^{(m)}_\mu&=&\frac{1}{\sqrt{2}}\left(W^{(m)+}_\mu \sigma^++W^{(m)-}_\mu \sigma^-\right)\nonumber \\
&&+\frac{1}{2c_W}Z^{(m)}_\mu \left(\sigma^3-2s^2_WQ\right)+s_WA^{(m)}_\mu Q\, , \ \ \ \ m=0,\, 1, \cdots \, ,
\end{eqnarray}
\begin{eqnarray}
{\cal O}^{(m)}_5&=&\frac{1}{\sqrt{2}}\left[\left(-s_\alpha H^{(m)+}+c_\alpha G^+_{W^{(m)}}\right)\sigma^++\left(s_\alpha H^{(m)-}-c_\alpha G^-_{W^{(m)}}\right)\sigma^-\right]\nonumber \\
&&+\frac{1}{2c_W}\left(-s_\beta A^{(m)}+c_\beta G_{Z^{(m)}}\right)\left(\sigma^3-2s^2_WQ\right)+s_WG_{A^{(m)}}Q\, , \, \, \, \, m=1,2,\cdots \, .
\end{eqnarray}
In the above expressions, $\sigma^\pm=\frac{1}{2}(\sigma^1\pm i\sigma^2)$ and $Q=\frac{\sigma^3}{2}+\frac{Y}{2}$. Also, it should be noticed that ${\cal O}^{(m)5}=-{\cal O}^{(m)}_5$. On the other hand, the Yang--Mills tensor structures appearing in Eq. (\ref{KT}) are given by
\begin{eqnarray}
{\cal W}^{(n)+}_{\mu 5}&=&\frac{n}{R}W^{(n)+}_\mu-is_\alpha \left(D^{(0)}_{e\mu}H^{(n)+}\right)+ic_\alpha\left(D^{(0)}_{e\mu}G^+_{W^{(n)}}\right)\nonumber \\
&&-gc_WZ^{(0)}_\mu\left(s_\alpha H^{(n)+}-c_\alpha G^+_{W^{(n)}}\right)\nonumber \\
&&+igW^{(0)+}_\mu \left[c_W\left(-s_\beta A^{(n)}+c_\beta G_{Z^{(n)}}\right)+s_WG_{A^{(n)}}\right]\nonumber \\
&&+ig\Delta'^{nrs}\Big\{W^{(r)+}_\mu \left[c_W\left(-s_\beta A^{(s)}+c_\beta G_{Z^{(s)}}\right)+s_WG_{A^{(s)}}\right]\nonumber \\
&&-i\left(c_WZ^{(r)}_\mu +s_W A^{(r)}_\mu\right)\left(s_\alpha H^{(s)+}-c_\alpha G^+_{W^{(s)}}\right)\Big\}\, ,
\end{eqnarray}
\begin{eqnarray}
{\cal W}^{(n)3}_{\mu 5}&=&\frac{n}{R}\left(c_WZ^{(n)}_\mu+s_WA^{(n)}_\mu\right)-c_W\left(s_\beta \partial_\mu A^{(n)}-c_\beta \partial_\mu G_{Z^{(n)}}\right)+s_W\partial_\mu G_{A^{(n)}}\nonumber \\
&&+g\left[s_\alpha \left(W^{(0)+}_\mu H^{(n)-}+W^{(0)-}_\mu H^{(n)+}\right)-c_\alpha \left(W^{(0)+}_\mu G^-_{W^{(n)}}+W^{(0)-}_\mu G^+_{W^{(n)}}\right)\right]\nonumber \\
&&+g\Delta'^{nrs}\left[s_\alpha \left(W^{(r)+}_\mu H^{(s)-}+W^{(r)-}_\mu H^{(s)+}\right)-c_\alpha \left(W^{(r)+}_\mu G^-_{W^{(s)}}+W^{(r)-}_\mu G^+_{W^{(s)}}\right)\right]\, ,
\end{eqnarray}
\begin{equation}
B^{(n)}_{\mu 5}=\frac{n}{R}\left(c_WA^{(n)}_\mu-s_WZ^{(n)}_\mu\right)+s_W\left(s_\beta \partial_\mu A^{(n)}-c_\beta \partial_\mu G_{Z^{(n)}}\right)+c_W\partial_\mu G_{A^{(n)}}\,.
\end{equation}
In the above expressions, $G_{W^{(n)}}$, $G_{Z^{(n)}}$, and $G_{A^{(n)}}$ are the pseudo Goldstone bosons associated with the $W^{(n)}$, $Z^{(n)}$, and $A^{(n)}$ KK gauge bosons, respectively (see \ref{bm} for their definition through the $\alpha$ and $\beta$ angles).

Many physical couplings are induced by the Lagrangian (\ref{KT}), but not all contribute at the one--loop level to SM observables. In particular, those terms that are proportional to the factors $\Delta^{nrs}$ or $\Delta'^{nrs}$ first contribute to SM observables at the two--loop level, although they generate contributions to non--SM observables still at the tree level. Such interactions involve only one SM field or no SM fields. Here, we derive explicit expressions for the pieces of ${\cal L}_{Kinetic}$ that generate the most important effects of the fifth dimension on SM Green's functions, which first occur at the one--loop level.

The couplings of the Higgs boson to pairs of KK excitations are given by
\begin{eqnarray}
{\cal L}_{H^{(0)}({\rm KK})({\rm KK})}&=&gm_{W^{(0)}}H^{(0)}\left(W^{(0)-}_\mu W^{(0)+\mu}+W^{(n)-}_\mu W^{(n)+\mu} \right)
\nonumber \\&&
+\frac{gm_{Z^{(0)}}}{2c_W}H^{(0)}\left(Z^{(0)}_\mu Z^{(0)\mu}+Z^{(n)}_\mu Z^{(n)\mu} \right)\nonumber \\
&&+\frac{igc_\alpha}{2}\Big[H^{(0)}\left(W^{(n)-}_\mu \partial^\mu H^{(n)+}-W^{(n)+}_\mu \partial^\mu H^{(n)-}\right)
\nonumber \\&&
-\left(W^{(n)-}_\mu H^{(n)+}-W^{(n)+}_\mu H^{(n)-}\right)\partial^\mu H^{(0)}\Big]\nonumber \\
&&-\frac{gc_\beta}{2c_W}Z^{(n)}_\mu \left(A^{(n)} \partial^\mu H^{(0)}-H^{(0)}\partial^\mu A^{(n)}\right)\, .
\end{eqnarray}
Notice that the standard couplings involving only zero modes have also been included. On the other hand, the electroweak gauge bosons couple to pairs of KK excitations as follows:
\begin{eqnarray}
{\cal L}_{W^{(0)}({\rm KK})({\rm KK})}&=&gm_{W^{(0)}}H^{(n)}\left(W^{(0)+}_\mu W^{(n)-\mu}+W^{(0)-}_\mu W^{(n)+\mu}\right) \nonumber \\
&&+\frac{g}{2}\left(2c_Ws_\alpha s_\beta+c_\alpha c_\beta \right)
\Big[A^{(n)}\left(W^{(0)+}_\mu \partial^\mu H^{(n)-}+W^{(0)-}_\mu \partial^\mu H^{(n)+}\right)\nonumber \\
&&-\left(W^{(0)+}_\mu H^{(n)-}+W^{(0)-}_\mu H^{(n)+}\right)\partial^\mu A^{(n)}\Big]\nonumber \\
&&-\frac{igc_\alpha}{2}\Big[H^{(n)}\left(W^{(0)+}_\mu \partial^\mu H^{(n)-}-W^{(0)-}_\mu \partial^\mu H^{(n)+}\right)\nonumber \\
&&-\left(W^{(0)+}_\mu H^{(n)-}-W^{(0)-}_\mu H^{(n)+}\right)\partial^\mu H^{(n)}\Big]\nonumber \\
&&+\frac{3}{2}igc_Wc_\beta m_{Z^{(0)}}A^{(n)}\left(W^{(0)-\mu}W^{(n)+}_\mu-W^{(0)+\mu}W^{(n)-}_\mu\right)\nonumber \\
&&+gc_\alpha \left[m_{W^{(0)}}s_WA^{(n)}_\mu-m_{Z^{(0)}}\left(\frac{1+2s^2_W}{4}\right)Z^{(n)}_\mu \right]
\nonumber \\&&\times
\left(W^{(0)+}_\mu H^{(n)-}+W^{(0)-}_\mu H^{(n)+}\right)\, ,
\end{eqnarray}
\begin{eqnarray}
{\cal L}_{Z^{(0)}({\rm KK})({\rm KK})}&=&-igc_W\left(1-\frac{c^2_\alpha}{2c^2_W}\right) Z^{(0)}_\mu \left(H^{(n)+}\partial^\mu H^{(n)-}-H^{(n)-}\partial^\mu H^{(n)+}\right)\nonumber \\
&&-gc_\alpha m_{Z^{(0)}}Z^{(0)}_\mu \left(H^{(n)-}W^{(n)+\mu}+H^{(n)+}W^{(n)-\mu}\right)\nonumber \\
&&+\frac{gc_\beta}{2c_W}Z^{(0)}_\mu \left(A^{(n)}\partial^\mu H^{(n)}-H^{(n)}\partial^\mu A^{(n)}\right)
+\frac{gm_{Z^{(0)}}}{c_W}Z^{(0)}_\mu Z^{(n)\mu}H^{(n)} \, ,
\end{eqnarray}
\begin{equation}
{\cal L}_{A^{(0)}({\rm KK})({\rm KK})}=-ieA^{(0)}_\mu \left(H^{(n)+}\partial^\mu H^{(n)-}-H^{(n)-}\partial^\mu H^{(n)+}\right) \, .
\end{equation}
The couplings of two Higgs bosons with pairs of KK excitations are given by
\begin{eqnarray}
{\cal L}_{H^{(0)}H^{(0)}({\rm KK})({\rm KK})}&=&\frac{g^2}{4}\,H^{(0)}H^{(0)}\left(W^{(0)-}_\mu W^{(0)+\mu}+W^{(n)-}_\mu W^{(n)+\mu}\right)\nonumber \\
&&+\frac{g^2}{8c^2_W}\,H^{(0)}H^{(0)}\left(Z^{(0)}_\mu Z^{(0)\mu}+Z^{(n)}_\mu Z^{(n)\mu}\right)\, .
\end{eqnarray}
There are also couplings involving a Higgs boson and an electroweak gauge boson with pairs of KK excitations:
\begin{eqnarray}
{\cal L}_{H^{(0)}V^{(0)}({\rm KK})({\rm KK})}&=&\frac{g^2}{2}\, H^{(0)}\Big[2c_\alpha\left(s_WA^{(0)\mu}-\frac{s^2_W}{c^2_W}Z^{(0)\mu}\right)\left(H^{(n)-}W^{(n)+}_\mu+H^{(n)+}W^{(n)-}_\mu\right)
\nonumber \\ &&
+\frac{1}{c^2_W}
H^{(n)}Z^{(0)}_\mu Z^{(n)\mu}+H^{(n)}\left(W^{(0)-\mu}W^{(n)+}_\mu+W^{(0)+\mu}W^{(n)-}_\mu\right)
\nonumber \\ &&
+ic_\beta A^{(n)}\left(W^{(0)-\mu}W^{(n)+}_\mu-W^{(0)+\mu}W^{(n)-}_\mu\right)\Big]\, .
\end{eqnarray}
Finally, the couplings of two gauge bosons with pairs of KK excitations are given by
\begin{eqnarray}
{\cal L}_{W^{(0)}W^{(0)}({\rm KK})({\rm KK})}&=&\frac{g^2}{4}\Big\{W^{(0)-}_\mu W^{(0)+\mu}\Big[2\left(1+s^2_\alpha \right)H^{(n)-} H^{(n)+}
\nonumber \\ &&
+H^{(n)}H^{(n)} +\left(c^2_\beta+4c^2_Ws^2_\beta \right)A^{(n)}A^{(n)} \Big]\nonumber \\
&&+2s^2_\alpha \left(W^{(0)+}_\mu W^{(0)+\mu}H^{(n)-}H^{(n)-}
+W^{(0)-}_\mu
W^{(0)-\mu}H^{(n)+}H^{(n)+} \right) \Big\}\, ,
\end{eqnarray}
\begin{eqnarray}
{\cal L}_{Z^{(0)}Z^{(0)}({\rm KK})({\rm KK})}&=&g^2Z^{(0)}_\mu Z^{(0)\mu}\Bigg[\left(c^2_Ws^2_\alpha+\frac{c_{2W}}{2c^2_W}c^2_\alpha\right)H^{(n)-}H^{(n)+}
\nonumber \\&&
+\frac{1}{4c^2_W}\left(H^{(n)}H^{(n)}
+c^2_\beta A^{(n)}A^{(n)} \right)\Bigg]\, ,
\end{eqnarray}
\begin{equation}
{\cal L}_{A^{(0)}A^{(0)}({\rm KK})({\rm KK})}=e^2\, A^{(0)}_\mu A^{(0)\mu}H^{(n)-}H^{(n)+}\, ,
\end{equation}
\begin{equation}
{\cal L}_{A^{(0)}Z^{(0)}({\rm KK})({\rm KK})}=egc_W\left(2s^2_\alpha+\frac{c_{2W}}{c^2_W}c^2_\alpha\right)A^{(0)}_\mu Z^{(0)\mu}H^{(n)-}H^{(n)+}\, ,
\end{equation}
\begin{eqnarray}
{\cal L}_{V^{(0)}W^{(0)}({\rm KK})({\rm KK})}&=&ig^2\Big[s_Wc_W\left(s_\alpha s_\beta+\frac{c_\alpha c_\beta}{2c_W}\right)A^{(0)\mu}\nonumber \\
&&+c^2_W\left(s_\alpha s_\beta+\frac{s^2_W}{2c^3_W}c_\alpha c_\beta\right)Z^{(0)\mu} \Big]A^{(n)}\left(W^{(0)+}_\mu H^{(n)-}-W^{(0)-}_\mu H^{(n)+}\right)\nonumber \\
&&+\frac{egc_\alpha}{2}\left(A^{(0)\mu}-\frac{s_W}{2c_W}Z^{(0)\mu}\right)H^{(n)}\left(W^{(0)+}_\mu H^{(n)-}+W^{(0)-}_\mu H^{(n)+}\right)\, .
\end{eqnarray}

\subsection{Trilinear and quartic gauge boson couplings}
We now turn to present the couplings among the SM gauge bosons and their KK excitations. These couplings are produced by the Yang--Mills sector and can be modified in a nontrivial way if a nonlinear gauge--fixing procedure, as the one discussed above, is introduced. We present the interactions that arise from the electroweak theory, which have been partially discussed in Ref. \cite{FMNRT}, in the context of a nonlinear gauge. We start by displaying those couplings that contribute to SM Green's functions at the one--loop level. In each case, the SM couplings will be included for comparison purposes. The trilinear vertices can be written as follows:
\begin{eqnarray}
{\cal L}_{W^{(0)3}W^{(n)-}W^{(n)+}}&=&-ig\Big[\left(W^{(0)+}_{\mu \nu}W^{(0)-\nu} -W^{(0)-}_{\mu \nu}W^{(0)+\nu} \right) W^{(0)3\mu}
\nonumber \\&&
+W^{(0)3}_{\mu \nu}W^{(0)-\mu}W^{(0)+\nu}+{\cal L}^{\rm SGT}_{\rm GF-1}\nonumber \\
&&+ \left(W^{(n)+}_{\mu \nu}W^{(n)-\nu} -W^{(n)-}_{\mu \nu}W^{(n)+\nu} \right) W^{(0)3\mu}
\nonumber \\&&
+W^{(0)3}_{\mu \nu}W^{(n)-\mu}W^{(n)+\nu}+{\cal L}^{\rm NSGT}_{\rm GF-1}
\Big]\, ,
\end{eqnarray}
where the gauge--dependent terms ${\cal L}^{\rm SGT}_{\rm GF-1}$ and ${\cal L}^{\rm NSGT}_{\rm GF-1}$ are present only if a nonlinear gauge--fixing procedure for removing the degeneration associated with the SGT~\cite{HT} and the NSGT~\cite{NT,FMNRT} has been introduced. Such terms do not exist in the unitary or linear $R_\xi$--gauge. In this expression, $W^{(0,n)3}=c_WZ^{(0,n)}+s_WA^{(0,n)}$, $W^{(0,n)\pm}_{\mu \nu}=\partial_\mu W^{(0,n)\pm}_\nu-\partial_\nu W^{(0,n)\pm}_\mu$, and $W^{(0,n)3}_{\mu \nu}=\partial_\mu W^{(0,n)3}_\nu-\partial_\nu W^{(0,n)3}_\mu$. It is important to notice that the Lorentz structure of the vertex $W^{(n)-}W^{(n)+}W^{(0)3}$ coincides exactly with that of $W^{(0)-}W^{(0)+}W^{(0)3}$, which is the SM vertex. On the other hand, there are also vertices containing only one SM $W^{(0)\pm}$ field, and they are given by the following Lagrangian:
\begin{eqnarray}
{\cal L}_{W^{(0)\mp}W^{(n)\pm}W^{(n)3}}&=&-ig\Big[\left(W^{(n)-}_{\mu \nu}W^{(n)3\nu}-W^{(n)3}_{\mu \nu}W^{(n)-\nu}  \right)W^{(0)+\mu}
+W^{(0)+}_{\mu \nu}W^{(n)3\mu}W^{(n)-\nu}\nonumber \\
&&-\left(W^{(n)+}_{\mu \nu}W^{(n)3\nu}-W^{(n)3}_{\mu \nu}W^{(n)+\nu}  \right)W^{(0)-\mu}
\nonumber \\&&
-W^{(0)-}_{\mu \nu}W^{(n)3\mu}W^{(n)+\nu} +{\cal L}^{\rm NSGT}_{\rm GF-2}
\Big]\, .
\end{eqnarray}

We can write the quartic vertices as
\begin{eqnarray}
{\cal L}_{W^{(0)3}W^{(0)3}W^{(n)-}W^{(n)+}}&=&\frac{g^2}{2}\Big[\left(W^{(0)-}_\mu W^{(0)3}_\nu-W^{(0)-}_\nu W^{(0)3}_\mu  \right)
\nonumber \\&&\times
\left(
W^{(0)+\nu}W^{(0)3\mu}-W^{(0)+\mu}W^{(0)3\nu}\right)
+{\cal L}^{\rm SGT}_{\rm GF-3}
\nonumber \\
&&
+\left(W^{(n)-}_\mu W^{(0)3}_\nu-W^{(n)-}_\nu W^{(0)3}_\mu  \right)
\nonumber \\&&\times
\left(
W^{(n)+\nu}W^{(0)3\mu}-W^{(n)+\mu}W^{(0)3\nu}\right)
+{\cal L}^{\rm NSGT}_{\rm GF-3}
\Big]\, .
\end{eqnarray}
Once again, notice that the $W^{(0)3}W^{(0)3}W^{(0)-}W^{(0)+}$ and $W^{(0)3}W^{(0)3}W^{(n)-}W^{(n)+}$ vertices have the same Lorentz structure. The above couplings involve only SM neutral gauge fields. The quartic vertices involving the SM charged fields are given by
\begin{eqnarray}
{\cal L}_{W^{(0)-}W^{(0)+}W^{(n)3}W^{(n)3}}&=&\frac{g^2}{2}\left(W^{(0)+}_\mu W^{(n)3}_\nu-W^{(n)3}_\mu W^{(0)+}_\nu \right)
\nonumber \\&&\times
\left(W^{(0)-\nu}W^{(n)3\mu}-W^{(0)-\mu}W^{(n)3\nu} \right)
+{\cal L}^{\rm NSGT}_{\rm GF-4} \, .
\end{eqnarray}
There are also couplings involving simultaneously one neutral and one charged SM field:
\begin{eqnarray}
{\cal L}_{W^{(0)\mp}W^{(n)\pm}W^{(0)3}W^{(n)3}}&=&-\frac{g^2}{2}\Big[\left(W^{(n)-}_\mu W^{(0)3}_\nu -W^{(n)-}_\nu W^{(0)3}_\mu\right)\left(W^{(0)+\mu}W^{(n)3\nu}-W^{(0)+\nu}W^{(n)3\mu}\right)\nonumber \\
&&+\left(W^{(0)+}_\mu W^{(0)3}_\nu-W^{(0)+}_\nu W^{(0)3}_\mu \right)\left(W^{(n)-\mu}W^{(n)3\nu}-W^{(n)-\nu}W^{(n)3\mu} \right)\nonumber \\
&&+{\rm H.\, c.} +{\cal L}^{\rm NSGT}_{\rm GF-5}
\Big]\, .
\end{eqnarray}
An odd number of KK excitations can appear in combination with the $\Delta^{nrs}$ symbol. Such combinations lead to the following couplings involving only one SM field:
\begin{eqnarray}
{\cal L}_{W^{(0)}W^{(n)}W^{(r)}W^{(s)}}&=&g^2\Delta^{nrs}\Bigg\{ \bigg(W^{(0)+}_\mu W^{(n)-}_\nu -W^{(0)+}_\nu W^{(n)-}_\mu
\nonumber \\&&
-W^{(0)-}_\mu W^{(n)+}_\nu
+W^{(0)-}_\nu W^{(n)+}_\mu \bigg)W^{(r)+\mu}W^{(s)-\nu}\nonumber \\
&-&\bigg[\bigg(W^{(n)-}_\mu W^{(0)3}_\nu- W^{(n)-}_\nu W^{(0)3}_\mu
\nonumber \\&&
+W^{(0)-}_\mu W^{(n)3}_\nu -W^{(0)-}_\nu W^{(n)3}_\mu\bigg)W^{(r)+\mu}W^{(s)3\nu}
+{\rm H.\, c.}\bigg]
\Bigg\}\, .
\end{eqnarray}
These couplings are not modified by covariant nonlinear gauge--fixing procedures as the one discussed above. These vertices only can contribute to SM Green's functions at two--loop or higher orders. There are no more couplings among electroweak gauge fields and their KK excitations. Although pure KK couplings first  contribute to SM observables at the two--loop level, they can contribute to nonstandard observables since lower levels. The corresponding trilinear and quartic vertices are given by the Lagrangian
\begin{eqnarray}
{\cal L}_{\rm KK-gauge}&=&-\frac{1}{2}\hat{W}^{(n)-}_{\mu \nu}\hat{W}^{(n)+\mu \nu}-\frac{1}{4}\hat{W}^{(n)3}_{\mu \nu}\hat{W}^{(n)3\mu \nu}\nonumber \\
&&-\frac{g^2}{2}\Big[W^{(m)+}_\mu W^{(m)-}_\nu \left(W^{(n)-\mu}W^{(n)+\nu}-W^{(n)+\mu}W^{(n)-\nu}\right)\nonumber \\
&&+2W^{(m)+}_\mu W^{(m)3}_\nu \left(W^{(n)-\mu}W^{(n)3\nu}-W^{(n)-\nu}W^{(n)3\mu}\right) \Big]\, ,
\end{eqnarray}
where
\begin{eqnarray}
\hat{W}^{(n)+}_{\mu \nu}&=&W^{(n)+}_{\mu \nu}+ig\Delta^{nrs}\left(W^{(r)+}_\mu W^{(s)3}_\nu -W^{(r)+}_\nu W^{(s)3}_\mu\right) \\ \nonumber \\
\hat{W}^{(n)3}_{\mu \nu}&=&W^{(n)3}_{\mu \nu}+ig\Delta^{nrs}\left(W^{(r)-}_\mu W^{(s)+}_\nu -W^{(r)+}_\nu W^{(s)-}_\mu\right) \, .
\end{eqnarray}
These trilinear and quartic vertices are not affected by the gauge fixing--procedure given above.

The couplings among gluons and their KK excitations have been discussed in Ref.~\cite{NT}, within the context of the nonlinear gauge--fixing procedure presented above.

\subsection{Scalar--fermion couplings}
The couplings of scalars with pairs of leptons or quarks arise from the Yukawa sector. To simplify the notation, we will use $f^{(0)}$, $f^{(n)}_a$ and $\hat{f}^{(n)}_a$ instead of $f'^{(0)}$, $\tilde{f}^{(n)}_a$ and $\tilde{\hat{f}}\hspace{0.001cm}^{(n)}_a$ for denoting the mass eigenstates fields (see \ref{fm}). For comparison purposes, in each case the SM vertex will be included, if present. The couplings of the Higgs boson $H^{(0)}$ to pairs of fermions are given by the Lagrangian
\begin{eqnarray}
{\cal L}_{H^{(0)}f^{(n)}f^{(n)}}&=&-\frac{gm_{f^{(0)}_a}}{2m_{W^{(0)}}}H^{(0)}\Big\{\bar{f}^{(0)}_af^{(0)}_a
+ \Big[\sin\alpha^{(n)}_{f_a} \left(\bar{f} ^{(n)}_a f^{(n)}_a+\bar{\hat{f}}\hspace{0.001cm}^{(n)}_a \hat{f}^{(n)}_a  \right)\nonumber \\&&
+\cos\alpha^{(n)}_{f_a} \left(\bar{f}^{(n)}_{a}\gamma_5 \hat{f}^{(n)}_{a} -\bar{\hat{f}}\hspace{0.001cm}^{(n)}_{a}\gamma_5 f^{(n)}_{a}\right)\Big]\Big\}\, ,
\end{eqnarray}
where $f_{a}$ stands for a charged lepton or quark and the $\alpha^{(n)}_{f_a}$ angle is defined in \ref{fm}. Notice the presence of flavor violating couplings. Other interactions of the type $\varphi^{(0)}\varphi^{(n)}\varphi^{(n)}$ are
\begin{eqnarray}
{\cal L}_{f^{(0)}f^{(n)}H^{(n)}}&=&-\frac{gm_{f^{(0)}_a}}{2m_{W^{(0)}}} \bar{f}^{(0)}_a\Bigg[\left(\sin\frac{\alpha^{(n)}_{f_a}}{2}P_R+\cos\frac{\alpha^{(n)}_{f_a}}{2}P_L \right)f^{(n)}_a\nonumber \\
&+&\left(\cos\frac{\alpha^{(n)}_{f_a}}{2}P_R+\sin\frac{\alpha^{(n)}_{f_a}}{2}P_L \right)\hat{f}^{(n)}_a\Bigg]H^{(n)}+{\rm H.\, c.}\, ,
\end{eqnarray}
\begin{eqnarray}
{\cal L}_{f^{(0)}f^{(n)}A^{(n)}}&=&-\frac{igm_{f^{(0)}_a}\, c_\beta}{2m_{W^{(0)}}}  \bar{f}^{(0)}_a
\Bigg[\left(\sin\frac{\alpha^{(n)}_{f_a}}{2}P_R-\cos\frac{\alpha^{(n)}_{f_a}}{2}P_L \right)f^{(n)}_a\nonumber \\
&+&\left(\cos\frac{\alpha^{(n)}_{f_a}}{2}P_R-\sin\frac{\alpha^{(n)}_{f_a}}{2}P_L \right)\hat{f}^{(n)}_a\Bigg]A^{(n)}+{\rm H.\, c.}
\end{eqnarray}
Regarding charged currents mediated by the $H^{(n)\pm}$ excitations, the lepton sector is flavor--family conserving, which is a consequence of the freedom to choice the unitary transformation (\ref{tn}). The corresponding Lagrangian is given by
\begin{equation}
{\cal L}_{\nu e H^+}=-\frac{gm_{e^{(0)}}c_\alpha}{m_{W^{(0)}}}\sum_{\rm families} \left(\bar{\nu}^{(0)}_L\hat{e}^{(n)}_R+
\bar{\nu}^{(n)}_L\hat{e}^{(0)}_R\right)H^{(n)+}+{\rm H.\, c.}
\end{equation}
In contrast, there is no way to avoid the presence of flavor violation in the quark sector. In this case, the Lagrangian characterizing the vertices of the way $q^{(0)}q^{(n)}H^{(n)\pm}$ is given by
\begin{eqnarray}
{\cal L}_{q^{(0)}q^{(n)}H^{(n)\pm}}&=&\frac{gc_\alpha}{\sqrt{2}m_{W^{(0)}}}H^{(n)+}
\Bigg[\bar{U}^{(0)}P_RV^{(0)}_d\left(\sin \frac{\alpha^{(n)}_d}{2} D^{(n)}+\cos \frac{\alpha^{(n)}_d}{2} \hat{D}^{(n)}\right)\nonumber \\
&&+\left(\cos \frac{\alpha^{(n)}_u}{2}\bar{U}^{(n)}+\sin \frac{\alpha^{(n)}_u}{2}\bar{\hat{U}}^{(n)}\right)P_RV^{(0)}_dD^{(0)} \Bigg]+\, {\rm H.\, c.} \nonumber \\
&&-\frac{gc_\alpha}{\sqrt{2}m_{W^{(0)}}}H^{(n)-}
\Bigg[\bar{D}^{(0)}P_RV^{(0)\dag}_u\left(\sin \frac{\alpha^{(n)}_u}{2} U^{(n)}+\cos \frac{\alpha^{(n)}_u}{2} \hat{U}^{(n)}\right)\nonumber \\
&&+\left(\cos \frac{\alpha^{(n)}_d}{2}\bar{D}^{(n)}+\sin \frac{\alpha^{(n)}_d}{2}\bar{\hat{D}}^{(n)}\right)P_RV^{(0)\dag}_uU^{(0)} \Bigg]
+\, {\rm H.\, c.}\, ,
\end{eqnarray}
where
\begin{eqnarray}
V^{(0)}_d&=&KM^{(0)}_d \, , \\
V^{(0)}_u&=&KM^{(0)}_u\, ,
\end{eqnarray}
with $K=V^u_LV^{d\dag}_L$ being the Cabibbo-Kobayashi-Maskawa matrix and $M^{(0)}_u (M^{(0)}_d)$ is the diagonal mass matrix of the SM quarks of type up (down).

The Yukawa sector also induces vertices involving only KK excitations. We first exhibit the charged currents mediated by the $H^{(n)\pm}$ scalars. In the lepton sector, these currents are given by
\begin{eqnarray}
{\cal L}_{H^{(s)+}N^{(n)}E^{(r)}}&=&-\frac{gc_\alpha}{\sqrt{2}m_{W^{(0)}}}H^{(s)+}
\Big[\sin \frac{\alpha^{(r)}_e}{2}\bar{N}^{(n)}M^{(0)}_e\left(\Delta^{nrs}P_R+\Delta'^{nrs}P_L\right)E^{(r)}\nonumber \\
&&+\cos \frac{\alpha^{(r)}_e}{2}\bar{N}^{(n)}M^{(0)}_e\left(\Delta^{nrs}P_R-\Delta'^{nrs}P_L\right)\hat{E}^{(r)} \Big]+\, {\rm H.\, c.}\, ,
\end{eqnarray}
where $M^{(0)}_e$ is the diagonal mass matrix of the SM charged leptons. On the other hand, in the quark sector such currents can be written as follows:

\begin{eqnarray}
{\cal L}_{H^{(s)\pm}U^{(n)}D^{(r)}}&=&-\frac{gc_\alpha}{\sqrt{2}m_{W^{(0)}}}H^{(s)+}
\Big[\sin \frac{\alpha^{(r)}_d}{2}\cos \frac{\alpha^{(n)}_u}{2} \bar{U}^{(n)}V^{(0)}_d\left(\Delta^{nrs}P_R+\Delta'^{nrs}P_L\right)D^{(r)}\nonumber \\
 &&+\sin \frac{\alpha^{(r)}_u}{2}\cos \frac{\alpha^{(n)}_d}{2} \bar{\hat{U}}^{(n)}V^{(0)}_d\left(\Delta^{nrs}P_R+\Delta'^{nrs}P_L\right)\hat{D}^{(r)}\nonumber \\
 &&+\cos \frac{\alpha^{(r)}_d}{2}\cos \frac{\alpha^{(n)}_u}{2}\bar{U}^{(n)}V^{(0)}_d\left(\Delta^{nrs}P_R-\Delta'^{nrs}P_L\right)\hat{D}^{(r)}\nonumber \\
 &&+\sin \frac{\alpha^{(r)}_d}{2}\sin \frac{\alpha^{(n)}_u}{2}\bar{\hat{U}}^{(n)}V^{(0)}_d\left(\Delta^{nrs}P_R-\Delta'^{nrs}P_L\right)D^{(r)}  \Big]
+\, {\rm H.\, c.}\nonumber \\ \nonumber \\ &&+\frac{gc_\alpha}{\sqrt{2}m_{W^{(0)}}}H^{(s)-}
\Big[\sin \frac{\alpha^{(r)}_u}{2}\cos \frac{\alpha^{(n)}_d}{2} \bar{D}^{(n)}V^{(0)}_u\left(\Delta^{nrs}P_R+\Delta'^{nrs}P_L\right)U^{(r)}\nonumber \\
 &&+\sin \frac{\alpha^{(r)}_d}{2}\cos \frac{\alpha^{(n)}_u}{2} \bar{\hat{D}}^{(n)}V^{(0)}_u\left(\Delta^{nrs}P_R+\Delta'^{nrs}P_L\right)\hat{U}^{(r)}\nonumber \\
 &&+\cos \frac{\alpha^{(r)}_u}{2}\cos \frac{\alpha^{(n)}_d}{2}\bar{D}^{(n)}V^{(0)}_u\left(\Delta^{nrs}P_R-\Delta'^{nrs}P_L\right)\hat{U}^{(r)}\nonumber \\
 &&+\sin \frac{\alpha^{(r)}_u}{2}\sin \frac{\alpha^{(n)}_d}{2}\bar{\hat{D}}^{(n)}V^{(0)}_u\left(\Delta^{nrs}P_R-\Delta'^{nrs}P_L\right)U^{(r)}  \Big]
+\, {\rm H.\, c.}
\end{eqnarray}
Finally, the neutral currents mediated by the $H^{(s)}$ and $A^{(s)}$ scalars can be written as follows:
\begin{eqnarray}
{\cal L}_{H^{(s)}\mathds{F}^{(n)}\mathds{F}^{(r)}}&=&-\frac{g}{2m_{W^{(0)}}}H^{(s)}\sum_{\mathds{F}=E,D,U}\Bigg\{\sin \frac{\alpha^{(n)}_\mathds{F}}{2} \cos \frac{\alpha^{(n)}_\mathds{F}}{2}\left[
\bar{\mathds{F}}^{(n)}M^{(0)}_{\mathds{F}}\left(\Delta^{nrs}+\Delta'^{nrs}\right)\mathds{F}^{(r)}
\right.
\nonumber \\ &&
\left.
+\bar{\hat{\mathds{F}}}^{(n)}M^{(0)}_{\mathds{F}}\left(\Delta^{nrs}+\Delta'^{nrs}\right)\hat{\mathds{F}}^{(r)}\right]\nonumber \\
&&+\bar{\mathds{F}}^{(n)}M^{(0)}_{\mathds{F}}\left[\Delta^{nrs}\left(\cos^2 \frac{\alpha^{(n)}_\mathds{F}}{2}P_R+\sin^2 \frac{\alpha^{(n)}_\mathds{F}}{2}P_L\right)
\right.
\nonumber \\&&
\left.
-\Delta'^{nrs}\left(\cos^2 \frac{\alpha^{(n)}_\mathds{F}}{2}P_L+\sin^2 \frac{\alpha^{(n)}_\mathds{F}}{2}P_R\right)\right]\hat{\mathds{F}}^{(r)}\nonumber \\
&&+\bar{\hat{\mathds{F}}}^{(n)}M^{(0)}_{\mathds{F}}\left[\Delta^{nrs}\left(\cos^2 \frac{\alpha^{(n)}_\mathds{F}}{2}P_L+\sin^2 \frac{\alpha^{(n)}_\mathds{F}}{2}P_R\right)
\right.
\nonumber \\ &&
\left.
-\Delta'^{nrs}\left(\cos^2 \frac{\alpha^{(n)}_\mathds{F}}{2}P_R+\sin^2 \frac{\alpha^{(n)}_\mathds{F}}{2}P_L\right)\right]\mathds{F}^{(r)} \Bigg\}\, ,
\end{eqnarray}

\begin{eqnarray}
{\cal L}_{A^{(s)}\mathds{F}^{(n)}\mathds{F}^{(r)}}&=&-\frac{ig}{2m_{W^{(0)}}}A^{(s)}\sum_{\mathds{F}=E,D,U}\Bigg\{\sin \frac{\alpha^{(n)}_\mathds{F}}{2} \cos \frac{\alpha^{(n)}_\mathds{F}}{2}\left[
\bar{\mathds{F}}^{(n)}M^{(0)}_\mathds{F}\left(\Delta^{nrs}-\Delta'^{nrs}\right)\gamma^5\mathds{F}^{(r)}
\right.
\nonumber \\ &&
\left.
+\bar{\hat{\mathds{F}}}^{(n)}M^{(0)}_\mathds{F}\left(\Delta^{nrs}-\Delta'^{nrs}\right)\gamma^5\hat{\mathds{F}}^{(r)}\right]\nonumber \\
&+&\bar{\mathds{F}}^{(n)}M^{(0)}_\mathds{F}\left[\Delta^{nrs}\left(\cos^2 \frac{\alpha^{(n)}_\mathds{F}}{2}P_R-\sin^2 \frac{\alpha^{(n)}_\mathds{F}}{2}P_L\right)
\right.
\nonumber \\ &&
\left.
-\Delta'^{nrs}\left(\cos^2 \frac{\alpha^{(n)}_\mathds{F}}{2}P_L-\sin^2 \frac{\alpha^{(n)}_\mathds{F}}{2}P_R\right)\right]\hat{\mathds{F}}^{(r)}\nonumber \\
&-&\bar{\hat{\mathds{F}}}^{(n)} M^{(0)}_\mathds {F}\left[\Delta^{nrs}\left(\cos^2 \frac{\alpha^{(n)}_\mathds{F}}{2}P_L-\sin^2 \frac{\alpha^{(n)}_\mathds{F}}{2}P_R\right)
\right.
\nonumber \\&&
\left.
-\Delta'^{nrs}\left(\cos^2 \frac{\alpha^{(n)}_\mathds{F}}{2}P_R-\sin^2 \frac{\alpha^{(n)}_\mathds{F}}{2}P_L\right)\right] \mathds{F}^{(r)} \Bigg\}.
\end{eqnarray}
In the above expressions, the symbol $M^{(0)}_\mathds{F}$ represents the diagonal SM fermionic mass matrix.

\subsection{Gauge boson--fermion couplings}
 The couplings which we are interested in are induced by the Currents sector, discussed above. Once again, we will use $f^{(0)}$, $f^{(n)}$, and $\hat{f}^{(n)}$ to denote the mass eigenstates fields instead of $f'^{(0)}$, $\tilde{f}^{(n)}$, and $\tilde{\hat{f}}\hspace{0.001cm}^{(n)}$. We first discuss the lepton sector, in which the couplings of the type $V^{(0)}f^{(n)}f^{(n)}$ are given by the following Lagrangians:
\begin{eqnarray}
{\cal L}_{W^{(0)}\nu^{(n)}e^{(n)}}&=&\frac{g}{\sqrt{2}} \left[\bar{N}^{(0)}_L\gamma^\mu E^{(0)}_L+\cos\frac{\alpha^{(n)}_e}{2}\left(\bar{N}^{(n)}\gamma^\mu E^{(n)}\right)
\right.
\nonumber \\&&
\left.
-\sin\frac{\alpha^{(n)}_e}{2}\left(\bar{N}^{(n)}\gamma^\mu \gamma^5 \hat{E}^{(n)}\right)\right]W^{(0)+}_\mu
+{\rm H.\, c.}\, ,
\end{eqnarray}
\begin{eqnarray}
{\cal L}_{Z^{(0)}e^{(n)}e^{(n)}}&=&\frac{g}{2c_W} \Bigg[\bar{N}^{(0)}_L\gamma^\mu N^{(0)}_L+\bar{E}^{(0)}\gamma^\mu \left(g^e_V-g^e_A\gamma^5\right)E^{(0)}+\bar{N}^{(n)}\gamma^\mu N^{(n)}\nonumber \\
&&\left(\bar{E}^{(n)} \, \, \, \bar{\hat{E}}^{(n)}\right)\gamma^\mu \left(\begin{array}{ccc}
Z_{EE} & Z_{E\hat{E}} \\
\, \\
Z_{\hat{E}E} & Z_{\hat{E}\hat{E}}
\end{array}\right) \left(\begin{array}{ccc}
E^{(n)} \\
\, \\
\hat{E}^{(n)}
\end{array}\right)\Bigg]Z^{(0)}_\mu \, ,
\end{eqnarray}
where
\begin{eqnarray}
Z_{EE}&=&\cos^2\frac{\alpha^{(n)}_e}{2}-2s^2_W \, , \\
Z_{\hat{E}\hat{E}}&=&\sin^2\frac{\alpha^{(n)}_e}{2}-2s^2_W \, , \\
Z_{E\hat{E}}&=&Z_{\hat{E}E}=\sin\frac{\alpha^{(n)}_e}{2}\cos\frac{\alpha^{(n)}_e}{2}\, .
\end{eqnarray}
On the other hand, the electromagnetic current is given by
\begin{equation}
{\cal L}_{A^{(0)}e^{(n)}e^{(n)}}=-e  \left(\bar{E}^{(0)}\gamma^\mu E^{(0)}+\bar{E}^{(n)}\gamma^\mu E^{(n)}+\bar{\hat{E}}^{(n)}\gamma^\mu \hat{E}^{(n)} \right)A^{(0)}_\mu \, .
\end{equation}
Notice that the SM couplings have been included. In this context, $g^e_V=-1/2+2s^2_W$ and $g^e_A=-1/2$. The couplings of the type $V^{(n)}f^{(0)}f^{(n)}$ are given by:
\begin{eqnarray}
{\cal L}_{W^{(n)}e^{(0)}e^{(n)}}&=&\frac{g}{\sqrt{2}} \left[\bar{N}^{(n)}_L\gamma^\mu E^{(0)}_L+\cos\frac{\alpha^{(n)}_e}{2}\left(\bar{N}^{(0)}_L\gamma^\mu E^{(n)}_L\right)
\right.
\nonumber \\&&
\left.
+\sin\frac{\alpha^{(n)}_e}{2}\left(\bar{N}^{(0)}_L\gamma^\mu \hat{E}^{(n)}_L\right)  \right]W^{(n)+}_\mu +{\rm H.\, c.} \, ,
\end{eqnarray}
\begin{eqnarray}
{\cal L}_{Z^{(n)}e^{(0)}e^{(n)}}&=&\frac{g}{2c_W} \Bigg[\bar{N}^{(0)}_L\gamma^\mu N^{(n)}_L+\bar{E}^{(0)}\gamma^\mu \left(
\left(2s^2_W-1\right)\cos\frac{\alpha^{(n)}_e}{2}P_L
-2s^2_W \sin\frac{\alpha^{(n)}_e}{2}P_R \right)E^{(n)}
\nonumber \\&&
+ \bar{E}^{(0)}\gamma^\mu \left(
\left(2s^2_W-1\right)\sin\frac{\alpha^{(n)}_e}{2}P_L
-2s^2_W \cos\frac{\alpha^{(n)}_e}{2}P_R \right)\hat{E}^{(n)} \Bigg]Z^{(n)}_\mu +{\rm H.\, c.}\, ,
\end{eqnarray}
\begin{eqnarray}
{\cal L}_{A^{(n)}e^{(0)}e^{(n)}}&=&-e \Bigg[\bar{E}^{(0)}\gamma^\mu \left(\cos\frac{\alpha^{(n)}_e}{2} P_L+ \sin\frac{\alpha^{(n)}_e}{2}P_R\right)E^{(n)}\nonumber \\
 &&+ \bar{E}^{(0)}\gamma^\mu \left(\cos\frac{\alpha^{(n)}_e}{2} P_R+ \sin\frac{\alpha^{(n)}_e}{2}P_L\right)\hat{E}^{(n)}\Bigg]A^{(n)}_\mu
+{\rm H.\, c.}
\end{eqnarray}

We now turn to discuss the couplings of the SM gauge bosons ($V^{(0)}=G^{(0)},W^{(0)}, Z^{(0)}, A^{(0)}$) with pairs of KK quark excitations. In the electroweak sector, the charged currents are given by the following Lagrangian
\begin{eqnarray}
{\cal L}_{W^{(0)}u^{(n)}d^{(n)}}&=&\frac{g}{\sqrt{2}}\Bigg[\bar{U}^{(0)}_LK\gamma^\mu D^{(0)}_L
\nonumber \\&&
+
\left(\bar{U}^{(n)} \, \, \, \bar{\hat{U}}^{(n)}\right)K\gamma^\mu \left(\begin{array}{ccc}
W_{UD} & W_{U\hat{D}} \\
\, \\
W_{\hat{U}D} & W_{\hat{U}\hat{D}}
\end{array}\right) \left(\begin{array}{ccc}
D^{(n)} \\
\, \\
\hat{D}^{(n)}
\end{array}\right)\Bigg]W^{(0)+}_\mu
+{\rm H.\, c.} \, ,
\end{eqnarray}
where
\begin{eqnarray}
W_{UD}&=&\cos \frac{\alpha^{(n)}_u}{2}\cos \frac{\alpha^{(n)}_d}{2} \, , \\
W_{\hat{U}\hat{D}}&=&\sin \frac{\alpha^{(n)}_u}{2}\sin \frac{\alpha^{(n)}_d}{2} \, , \\
W_{U\hat{D}}&=&-\sin \frac{\alpha^{(n)}_d}{2}\cos \frac{\alpha^{(n)}_u}{2}\, ,\\
W_{\hat{U}D}&=&-\sin \frac{\alpha^{(n)}_u}{2}\cos \frac{\alpha^{(n)}_d}{2}\, .
\end{eqnarray}
As far as the neutral currents are concerned, they have the following structure
\begin{eqnarray}
{\cal L}_{Z^{(0)}q^{(n)}q^{(n)}}&=&\frac{g}{c_W}Z^{(0)}_\mu\Bigg\{ \bar{U}^{(0)}\gamma^\mu \left( g^u_V-g^u_A\gamma^5\right)U^{(0)}+ \bar{D}^{(0)}\gamma^\mu \left( g^d_V-g^d_A\gamma^5\right)D^{(0)}\nonumber \\
&&+\left(\bar{U}^{(n)} \, \, \, \bar{\hat{U}}\hspace{0.001cm}^{(n)}\right)\gamma^\mu \left(\begin{array}{ccc}
Z^q_{UU} & Z^q_{U\hat{U}}\gamma^5 \\
\, \\
Z^q_{\hat{U}U}\gamma^5 & Z^q_{\hat{U}\hat{U}}
\end{array}\right) \left(\begin{array}{ccc}
U^{(n)} \\
\, \\
\hat{U}^{(n)}
\end{array}\right) \nonumber \\
&&+\left(\bar{D}^{(n)} \, \, \, \bar{\hat{D}}\hspace{0.001cm}^{(n)}\right)\gamma^\mu \left(\begin{array}{ccc}
Z^q_{DD} & Z^q_{D\hat{D}}\gamma^5 \\
\, \\
Z^q_{\hat{D}D}\gamma^5 & Z^q_{\hat{D}\hat{D}}
\end{array}\right) \left(\begin{array}{ccc}
D^{(n)} \\
\, \\
\hat{D}^{(n)}
\end{array}\right)
\Bigg\},
\end{eqnarray}
where
\begin{eqnarray}
Z^q_{UU}&=&\left( \frac{1}{2}-Q_u\right)\cos^2\frac{\alpha^{(n)}_{u}}{2}-Q_u s^2_W\sin^2\frac{\alpha^{(n)}_{u}}{2} \, , \\
Z^q_{\hat{U}\hat{U}}&=&\left( \frac{1}{2}-Q_u\right)\sin^2\frac{\alpha^{(n)}_{u}}{2}-Q_u s^2_W\cos^2\frac{\alpha^{(n)}_{u}}{2}\, , \\
Z^q_{U\hat{U}}&=&Z^q_{\hat{U}U}=-\left(\frac{1}{2}-Q_uc^2_W\right)\sin\frac{\alpha^{(n)}_{u}}{2}\cos\frac{\alpha^{(n)}_{u}}{2}\, ,
\end{eqnarray}
\begin{eqnarray}
Z^q_{DD}&=&\left( \frac{1}{2}+Q_d\right)\cos^2\frac{\alpha^{(n)}_{d}}{2}+Q_d s^2_W\sin^2\frac{\alpha^{(n)}_{d}}{2} \, , \\
Z^q_{\hat{D}\hat{D}}&=&\left( \frac{1}{2}+Q_d\right)\sin^2\frac{\alpha^{(n)}_{d}}{2}+Q_d s^2_W\cos^2\frac{\alpha^{(n)}_{d}}{2} \, , \\
Z^q_{D\hat{D}}&=&Z^q_{\hat{D}D}=-\left(\frac{1}{2}+Q_dc^2_W\right)\sin\frac{\alpha^{(n)}_{d}}{2}\cos\frac{\alpha^{(n)}_{d}}{2} \, .
\end{eqnarray}
In addition,  $g^{u,d}_V=1/2-Q_{u,d}s^2_W$ and $g^{u,d}_A=1/2$. The electromagnetic current is given by
\begin{equation}
{\cal L}_{A^{(0)}q^{(n)}q^{(n)}}=e\sum_{q=u,d,\ldots }Q_q\left(\bar{q}^{(0)}\gamma^\mu q^{(0)}+
\bar{q}^{(n)}\gamma^\mu q^{(n)}+\bar{\hat{q}}^{(n)}\gamma^\mu \hat{q}^{(n)}\right)A^{(0)}_\mu \,.
\end{equation}
The couplings of quarks to SM gluons are also diagonal:
\begin{eqnarray}
{\cal L}_{G^{(0)}q^{(n)}q^{(n)}}&=&g_{\rm s} \sum_{q=u,d,\ldots }\left(\bar{q}^{(0)}\gamma^\mu \frac{\lambda^a}{2}q^{(0)}
+
\bar{q}^{(n)}\gamma^\mu \frac{\lambda^a}{2}q^{(n)}+\bar{\hat{q}}^{(n)}\gamma^\mu \frac{\lambda^a}{2}\hat{q}^{(n)}\right)G^{(0)a}_\mu \,.
\end{eqnarray}

On the other hand, the couplings of a SM quark $q^{(0)}$ with pairs of KK excitations $q^{(n)}V^{(n)}$ ($V^{(n)}=G^{(n)a}, W^{(n)}, Z^{(n)}, A^{(n)}$) can be written as follows:
\begin{eqnarray}
{\cal L}_{W^{(n)}q^{(0)}q^{(n)}}&=&\frac{g}{\sqrt{2}}\Bigg\{ \bar{U}^{(0)}K\gamma^\mu \left[\left(\cos\frac{\alpha^{(n)}_{d}}{2}P_L+\sin\frac{\alpha^{(n)}_{d}}{2}P_R\right)D^{(n)}
\right.
\nonumber \\&&
\left.
+\left(\sin\frac{\alpha^{(n)}_{d}}{2}P_L+
\cos\frac{\alpha^{(n)}_{d}}{2}P_R\right)\hat{D}^{(n)}\right]W^{(n)+}_\mu \nonumber \\
&&+\bar{D}^{(0)}K^\dag \gamma^\mu \left[\left(\cos\frac{\alpha^{(n)}_{u}}{2}P_L+
\sin\frac{\alpha^{(n)}_{u}}{2}P_R\right)U^{(n)}
\right.
\nonumber \\&&
\left.
+\left(\sin\frac{\alpha^{(n)}_{u}}{2}P_L+
\cos\frac{\alpha^{(n)}_{u}}{2}P_R\right)\hat{U}^{(n)}\right]W^{(n)-}_\mu
\Bigg\}+{\rm H.\, c.}
\end{eqnarray}
\begin{eqnarray}
{\cal L}_{Z^{(n)}q^{(0)}q^{(n)}}&=&\frac{g}{c_W}\Bigg\{ \left(\frac{1}{2}-s^2_WQ_u\right)\bar{U}^{(0)}\gamma^\mu \left[\left(\cos\frac{\alpha^{(n)}_{u}}{2}P_L+\sin\frac{\alpha^{(n)}_{u}}{2}P_R\right)U^{(n)}
\right.
\nonumber \\ &&
\left.
+\left(\sin\frac{\alpha^{(n)}_{u}}{2}P_L+
\cos\frac{\alpha^{(n)}_{u}}{2}P_R\right)\hat{U}^{(n)}\right] \nonumber \\
&&+\left(\frac{1}{2}+s^2_WQ_d\right)\bar{D}^{(0)} \gamma^\mu \left[\left(\cos\frac{\alpha^{(n)}_{d}}{2}P_L+\sin\frac{\alpha^{(n)}_{d}}{2}P_R\right)D^{(n)}
\right.
\nonumber \\ &&
\left.
+\left(\sin\frac{\alpha^{(n)}_{d}}{2}P_L+
\cos\frac{\alpha^{(n)}_{d}}{2}P_R\right)\hat{D}^{(n)}\right]
\Bigg\}Z^{(n)}_\mu +{\rm H. \, c.}\, ,
\end{eqnarray}

\begin{eqnarray}
{\cal L}_{A^{(n)}q^{(0)}q^{(n)}}&=&e\Bigg\{Q_u\bar{U}^{(0)}\gamma^\mu \left[\left(\cos\frac{\alpha^{(n)}_{u}}{2}P_L+\sin\frac{\alpha^{(n)}_{u}}{2}P_R\right)U^{(n)}
\right.
\nonumber \\&&
\left.
+\left(\sin\frac{\alpha^{(n)}_{u}}{2}P_L+
\cos\frac{\alpha^{(n)}_{u}}{2}P_R\right)\hat{U}^{(n)}\right] \nonumber \\
&&+Q_d\bar{D}^{(0)} \gamma^\mu \left[\left(\cos\frac{\alpha^{(n)}_{d}}{2}P_L+\sin\frac{\alpha^{(n)}_{d}}{2}P_R\right)D^{(n)}
\right.
\nonumber \\&&
\left.
+\left(\sin\frac{\alpha^{(n)}_{d}}{2}P_L+
\cos\frac{\alpha^{(n)}_{d}}{2}P_R\right)\hat{D}^{(n)}\right]
\Bigg\}A^{(n)}_\mu
+{\rm H. \, c.} \, ,
\end{eqnarray}

\begin{eqnarray}
{\cal L}_{G^{(n)}q^{(0)}q^{(n)}}&=&g_s\Bigg\{\bar{U}^{(0)}\gamma^\mu \frac{\lambda^a}{2} \left[\left(\cos\frac{\alpha^{(n)}_{u}}{2}P_L+\sin\frac{\alpha^{(n)}_{u}}{2}P_R\right)U^{(n)}
\right.
\nonumber \\&&
\left.
+\left(\sin\frac{\alpha^{(n)}_{u}}{2}P_L+
\cos\frac{\alpha^{(n)}_{u}}{2}P_R\right)\hat{U}^{(n)}\right] \nonumber \\
&&+\bar{D}^{(0)} \gamma^\mu \frac{\lambda^a}{2} \left[\left(\cos\frac{\alpha^{(n)}_{d}}{2}P_L+\sin\frac{\alpha^{(n)}_{d}}{2}P_R\right)D^{(n)}
\right.
\nonumber \\&&
\left.
+\left(\sin\frac{\alpha^{(n)}_{d}}{2}P_L+
\cos\frac{\alpha^{(n)}_{d}}{2}P_R\right)\hat{D}^{(n)}\right]
\Bigg\}G^{(n)a}_\mu
+{\rm H. \, c.}
\end{eqnarray}

We now turn to discuss those couplings that involve only KK excitations. The corresponding couplings in the lepton sector are given by
\begin{eqnarray}
{\cal L}_{\nu^{(n)}e^{(s)}W^{(r)}}&=&\frac{g}{\sqrt{2}}\Big[\cos\frac{\alpha^{(s)}_{e}}{2}\,\bar{N}^{(n)}\gamma^\mu\left(\Delta^{nrs}P_L+\Delta'^{nrs}P_R\right)E^{(s)}
\nonumber \\
&&+\sin\frac{\alpha^{(s)}_{e}}{2}\,\bar{N}^{(n)}\gamma^\mu\left(\Delta^{nrs}P_L-\Delta'^{nrs}P_R\right)\hat{E}^{(s)}\Big]W^{(r)+}_\mu
+{\rm H.\, c.} \, ,
\end{eqnarray}
\begin{eqnarray}
{\cal L}_{e^{(n)}e^{(s)}Z^{(r)}}&=&\frac{g}{2c_W}Z^{(r)}_\mu\Bigg[\bar{N}^{(n)}\gamma^\mu \left(\Delta^{nrs}P_L+\Delta'^{nrs}P_R\right)N^{(s)} \nonumber \\
&&+\left(\bar{E}^{(n)} \, \, \, \bar{\hat{E}}^{(n)}\right)\gamma^\mu \left(\begin{array}{ccc}
Z^{nrs}_{EE} & Z^{nrs}_{E\hat{E}} \\
\, \\
Z^{nrs}_{\hat{E}E} & Z^{nrs}_{\hat{E}\hat{E}}
\end{array}\right) \left(\begin{array}{ccc}
E^{(s)} \\
\, \\
\hat{E}^{(s)}
\end{array}\right) \Bigg] \, ,
\end{eqnarray}
where
\begin{eqnarray}
Z^{nrs}_{EE}&=&\left[\cos^2\frac{\alpha^{(s)}_e}{2}\left(2s^2_W-1\right)\Delta^{nrs}+2\sin^2\frac{\alpha^{(s)_e}}{2}s^2_W \Delta'^{nrs}\right]P_L\nonumber \\
&&+\left[\cos^2\frac{\alpha^{(s)}_e}{2}\left(2s^2_W-1\right)\Delta'^{nrs}+2\sin^2\frac{\alpha^{(s)}_e}{2}s^2_W \Delta^{nrs}\right]P_R \, ,
\end{eqnarray}

\begin{eqnarray}
Z^{nrs}_{\hat{E}\hat{E}}&=&\left[\sin^2\frac{\alpha^{(s)}_e}{2}\left(2s^2_W-1\right)\Delta^{nrs}+2\cos^2\frac{\alpha^{(s)}_e}{2}s^2_W \Delta'^{nrs}\right]P_L\nonumber \\
&&+\left[\sin^2\frac{\alpha^{(s)}_e}{2}\left(2s^2_W-1\right)\Delta'^{nrs}+2\cos^2\frac{\alpha^{(s)}_e}{2}s^2_W \Delta^{nrs}\right]P_R \, ,
\end{eqnarray}

\begin{eqnarray}
Z^{nrs}_{E\hat{E}}=Z^{nrs}_{\hat{E}E}&=&\sin\frac{\alpha^{(s)}_e}{2}\cos\frac{\alpha^{(s)}_e}{2}\Big\{\left[\left(2s^2_W-1\right)\Delta^{nrs}-2s^2_W\Delta'^{rns}\right]P_L \nonumber \\
&&-\left[\left(2s^2_W-1\right)\Delta'^{nrs}-2s^2_W\Delta^{rns}\right]P_R\Big\} \, ,
\end{eqnarray}

\begin{equation}
{\cal L}_{e^{(n)}e^{(s)}A^{(r)}}=-e\left[\left(\bar{E}^{(n)} \, \, \, \bar{\hat{E}}^{(n)}\right)\gamma^\mu \left(\begin{array}{ccc}
A^{nrs}_{EE} & A^{nrs}_{E\hat{E}} \\
\, \\
A^{nrs}_{\hat{E}E} & A^{nrs}_{\hat{E}\hat{E}}
\end{array}\right) \left(\begin{array}{ccc}
E^{(s)} \\
\, \\
\hat{E}^{(s)}
\end{array}\right) \right]A^{(r)}_\mu \, ,
\end{equation}
with
\begin{eqnarray}
A^{nrs}_{EE}&=&\left(\cos^2\frac{\alpha^{(s)}_e}{2}\Delta^{nrs}+\sin^2\frac{\alpha^{(s)}_e}{2} \Delta'^{nrs}\right)P_L
\nonumber \\ &&
+\left(\cos^2\frac{\alpha^{(s)}_e}{2}\Delta'^{nrs}+\sin^2\frac{\alpha^{(s)}_e}{2} \Delta^{nrs}\right)P_R \, , \\
A^{nrs}_{\hat{E}\hat{E}}&=&\left(\sin^2\frac{\alpha^{(s)}_e}{2}\Delta^{nrs}+\cos^2\frac{\alpha^{(s)}_e}{2} \Delta'^{nrs}\right)P_L
\nonumber \\ &&
+\left(\sin^2\frac{\alpha^{(s)}_e}{2}\Delta'^{nrs}+\cos^2\frac{\alpha^{(s)}_e}{2} \Delta^{nrs}\right)P_R \, , \\
A^{nrs}_{E\hat{E}}&=&A^{nrs}_{\hat{E}E}=\sin\frac{\alpha^{(s)}_e}{2}\cos\frac{\alpha^{(s)}_e}{2}\left(\Delta^{nrs}-\Delta'^{nrs}\right) \, .
\end{eqnarray}

As far as the quark sector is concerned, the charged currents can be written as follows
\begin{eqnarray}
{\cal L}_{q^{(n)}q^{(s)}W^{(r)}}&=&\frac{g}{\sqrt{2}}\left[\left(\bar{U}^{(n)} \, \, \, \bar{\hat{U}}^{(n)}\right)K\gamma^\mu \left(\begin{array}{ccc}
W^{nrs}_{UD} & W^{nrs}_{U\hat{D}} \\
\, \\
W^{nrs}_{\hat{U}D} & W^{nrs}_{\hat{U}\hat{D}}
\end{array}\right) \left(\begin{array}{ccc}
D^{(s)} \\
\, \\
\hat{D}^{(s)}
\end{array}\right) \right]W^{(r)+}_\mu
+{\rm H. \, c.} \, ,
\end{eqnarray}
where
\begin{eqnarray}
W^{nrs}_{UD}&=&\cos\frac{\alpha^{(n)}_u}{2}\cos\frac{\alpha^{(s)}_d}{2}\left(\Delta^{nrs}P_L+\Delta'^{nrs}P_R\right) \\
W^{nrs}_{\hat{U}\hat{D}}&=&\sin\frac{\alpha^{(n)}_u}{2}\sin\frac{\alpha^{(s)}_d}{2}\left(\Delta^{nrs}P_L+\Delta'^{nrs}P_R\right) \\
W^{nrs}_{U\hat{D}}&=&\cos\frac{\alpha^{(n)}_u}{2}\sin\frac{\alpha^{(s)}_d}{2}\left(\Delta^{nrs}P_L-\Delta'^{nrs}P_R \right)\\
W^{nrs}_{\hat{U}D}&=&\sin\frac{\alpha^{(n)}_u}{2}\cos\frac{\alpha^{(s)}_d}{2}\left(\Delta^{nrs}P_L-\Delta'^{nrs}P_R \right)\, .
\end{eqnarray}
On the other hand, the neutral currents are given by
\begin{eqnarray}
&&{\cal L}_{q^{(n)}q^{(s)}Z^{(r)}}=\frac{g}{2c_W}\sin\frac{\alpha^{(s)}_q}{2}\cos\frac{\alpha^{(s)}_q}{2}Z^{(r)}_\mu
\nonumber \\&&\times
\Bigg[\left(\bar{U}^{(n)} \, \, \, \bar{\hat{U}}^{(n)}\right)\gamma^\mu \left(\begin{array}{ccc}
\cot\frac{\alpha^{(s)}_q}{2}Z^{nrs}_{UU} & Z^{nrs}_{U\hat{U}} \\
\, \\
Z^{nrs}_{\hat{U}U} & \tan\frac{\alpha^{(s)}_q}{2}Z^{nrs}_{\hat{U}\hat{U}}
\end{array}\right) \left(\begin{array}{ccc}
U^{(s)} \\
\, \\
\hat{U}^{(s)}
\end{array}\right)\nonumber \\
&&-\left(\bar{D}^{(n)} \, \, \, \bar{\hat{D}}^{(n)}\right)\gamma^\mu \left(\begin{array}{ccc}
\cot\frac{\alpha^{(s)}_q}{2}Z^{nrs}_{DD} & Z^{nrs}_{D\hat{D}} \\
\, \\
Z^{nrs}_{\hat{D}D} & \tan\frac{\alpha^{(s)}_q}{2}Z^{nrs}_{\hat{D}\hat{D}}
\end{array}\right) \left(\begin{array}{ccc}
D^{(s)} \\
\, \\
\hat{D}^{(s)}
\end{array}\right) \Bigg]
\end{eqnarray}
where
\begin{eqnarray}
Z^{nrs}_{UU}&=&Z^{nrs}_{\hat{U}\hat{U}}=\left[\left(1-2s^2_WQ_u\right)\Delta^{nrs}-2s^2_WQ_u \Delta'^{nrs}\right]P_L
\nonumber \\&&
+\left[\left(1-2s^2_WQ_u\right)\Delta'^{nrs}-2s^2_WQ_u \Delta^{nrs}\right]P_R \, , \\\nonumber \\
Z^{nrs}_{U\hat{U}}&=&Z^{nrs}_{\hat{U}U}=\left[\left(1-2s^2_WQ_u\right)\Delta^{nrs}-2s^2_WQ_u \Delta'^{nrs}\right]P_L
\nonumber \\&&
-\left[\left(1-2s^2_WQ_u\right)\Delta'^{nrs}-2s^2_WQ_u \Delta^{nrs}\right]P_R \, ,
\end{eqnarray}
\begin{eqnarray}
Z^{nrs}_{DD}&=&Z^{nrs}_{\hat{D}\hat{D}}=Z^{nrs}_{UU}(Q_u\to -Q_d) \, ,\\ \nonumber \\
Z^{nrs}_{D\hat{D}}&=&Z^{nrs}_{\hat{D}D}=Z^{nrs}_{U\hat{U}}(Q_u\to -Q_d)\, ,
\end{eqnarray}

\begin{eqnarray}
{\cal L}_{q^{(n)}q^{(s)}A^{(r)}}&=&eA^{(r)}_\mu\Bigg[Q_u\, \left(\bar{U}^{(n)} \, \, \, \bar{\hat{U}}^{(n)}\right)\gamma^\mu \left(\begin{array}{ccc}
A^{nrs}_{UU} & A^{nrs}_{U\hat{U}} \\
\, \\
A^{nrs}_{\hat{U}U} & A^{nrs}_{\hat{U}\hat{U}}
\end{array}\right) \left(\begin{array}{ccc}
U^{(s)} \\
\, \\
\hat{U}^{(s)}
\end{array}\right)\nonumber \\
&&+Q_d\, \left(\bar{D}^{(n)} \, \, \, \bar{\hat{D}}^{(n)}\right)\gamma^\mu \left(\begin{array}{ccc}
A^{nrs}_{DD} & A^{nrs}_{D\hat{D}} \\
\, \\
A^{nrs}_{\hat{D}D} & A^{nrs}_{\hat{D}\hat{D}}
\end{array}\right) \left(\begin{array}{ccc}
D^{(s)} \\
\, \\
\hat{D}^{(s)}
\end{array}\right) \Bigg]
\end{eqnarray}
with
\begin{eqnarray}
A^{nrs}_{UU}&=&A^{nrs}_{DD}=\left(\cos^2\frac{\alpha^{(s)}_{q}}{2}\Delta^{nrs}+\sin^2\frac{\alpha^{(s)}_{q}}{2}\Delta'^{nrs}\right)P_L
\nonumber \\&&
+\left(\sin^2\frac{\alpha^{(s)}_{q}}{2}\Delta^{nrs}+\cos^2\frac{\alpha^{(s)}_{q}}{2}\Delta'^{nrs}\right)P_R\, , \\
A^{nrs}_{\hat{U}\hat{U}}&=&A^{nrs}_{\hat{D}\hat{D}}=\left(\sin^2\frac{\alpha^{(s)}_{q}}{2}\Delta^{nrs}+\cos^2\frac{\alpha^{(s)}_{q}}{2}\Delta'^{nrs}\right)P_L
\nonumber \\&&
+\left(\cos^2\frac{\alpha^{(s)}_{q}}{2}\Delta^{nrs}+\sin^2\frac{\alpha^{(s)}_{q}}{2}\Delta'^{nrs}\right)P_R\, , \\
A^{nrs}_{U\hat{U}}&=&A^{nrs}_{\hat{U}U}=A^{nrs}_{D\hat{D}}=A^{nrs}_{\hat{D}D}=\sin\frac{\alpha^{(s)}_{q}}{2}\cos\frac{\alpha^{(s)}_{q}}{2}\left(\Delta^{nrs}
-\Delta'^{nrs}\right)\, .
\end{eqnarray}

Finally, the couplings to the gluon are given by
\begin{equation}
{\cal L}_{q^{(n)}q^{(s)}G^{(r)}}=g_s\sum_{q}\left[\left(\bar{q}^{(n)} \, \, \, \bar{\hat{q}}^{(n)}\right)\frac{\lambda^a}{2}\gamma^\mu \left(\begin{array}{ccc}
A^{nrs}_{qq} & Q^{nrs}_{q\hat{q}} \\
\, \\
Q^{nrs}_{\hat{q}q} & Q^{nrs}_{\hat{q}\hat{q}}
\end{array}\right) \left(\begin{array}{ccc}
q^{(s)} \\
\, \\
\hat{q}^{(s)}
\end{array}\right)\right]G^{(r)a}_\mu \, ,
\end{equation}
where
\begin{eqnarray}
Q^{nrs}_{qq}&=&\left(\cos^2\frac{\alpha^{(s)}_{q}}{2}\Delta^{nrs}+\sin^2\frac{\alpha^{(s)}_{q}}{2}\Delta'^{nrs}\right)P_L
\nonumber \\&&
+\left(\sin^2\frac{\alpha^{(s)}_{q}}{2}\Delta^{nrs}+\cos^2\frac{\alpha^{(s)}_{q}}{2}\Delta'^{nrs}\right)P_R \, ,\\
Q^{nrs}_{\hat{q}\hat{q}}&=&\left(\sin^2\frac{\alpha^{(s)}_{q}}{2}\Delta^{nrs}+\cos^2\frac{\alpha^{(s)}_{q}}{2}\Delta'^{nrs}\right)P_L
\nonumber \\&&
+\left(\cos^2\frac{\alpha^{(s)}_{q}}{2}\Delta^{nrs}+\sin^2\frac{\alpha^{(s)}_{q}}{2}\Delta'^{nrs}\right)P_R \, ,\\
Q^{nrs}_{q\hat{q}}&=&Q^{nrs}_{\hat{q}q}=\sin\frac{\alpha^{(s)}_{q}}{2}\cos\frac{\alpha^{(s)}_{q}}{2}\left(\Delta^{nrs}-\Delta'^{nrs}\right)\, .
\end{eqnarray}

\section{Summary}
\label{c}
In this paper, a comprehensive analysis of the SM in five dimensions, with the extra dimension compactified on the orbifold $S^1/Z_2$ of radius $R$, was presented. The mass eingenstate fields were determined in both the fermionic and bosonic sectors. The KK zero modes, which coincide with the SM fields, are endowed with mass through the usual Higgs mechanism, whereas the masses of the KK excitations receive contributions from both the compactification and the Higgs mechanism. The masses of the fermionic KK excitations receive contributions from the Currents sector via compactification and also from the Yukawa sector through the Higgs mechanism. It occurs that there is a double multiplicity, $f^{(n)}$ and $\hat{f}^{(n)}$, for each charged lepton and quark, which are mass degenerate, with mass given by $m^2_{f^{(n)}}=\left(\frac{n}{R}\right)^2+m^2_{f^{(0)}}$. The zero modes of the neutrinos remain massless, for they are the well known SM left--handed neutrinos, whereas their KK excitations arise in both types of helicities and have, therefore, a mass given by $m_{\nu^{(n)}}=\frac{n}{R}$. In contrast with the case of charged fermions, there are no partners of the $\nu^{(n)}$ neutrinos, as there are no right--handed neutrinos in the four--dimensional theory. For each species of charged fermions, a mix between it and its partner arises, which is characterized by an angle given by $\tan \alpha^{(n)}_{f_a}=\frac{m_{f^{(0)}}}{\frac{n}{R}}$. The SM gauge bosons receive their masses via the Higgs mechanism, whereas their KK excitations are endowed with mass through the compactification and the Higgs mechanisms. While the mass contribution produced by compactification is engendered in the Yang--Mills sector, that introduced by the Higgs mechanism comes from the Higgs kinetic term. These masses are given by $m^2_{V^{(n)}}=\left(\frac{n}{R}\right)^2+m^2_{V^{(0)}}$, for $V=W,Z$, and $m_{V^{(n)}}=\left(\frac{n}{R}\right)$, for $V=\gamma, g$. On the other hand, the KK excitations of the Higgs doublet, $\Phi^{(n)}$, determine physical KK excitations, namely, the charged $H^{(n)\pm}$ scalars, which have masses given by $m_{H^{(n)+}}=m_{W^{(n)+}}$, so they can be seen as KK excitations of the pseudo--Goldstone boson $G^\pm_{W^{(0)}}$ associated with the $W^{(0)\pm}$ gauge boson. With respect to the down component of $\Phi^{(n)}$, its real part, $H^{(n)}$, represents a KK excitation of the SM Higgs boson, $H^{(0)}$, and has a mass given by $m^2_{H^{(n)}}=\left(\frac{n}{R}\right)^2+m^2_{H^{(0)}}$. The imaginary part of this component, $A^{(n)}$, represents a neutral CP--odd scalar with mass $m_{A^{(n)}}=m_{Z^{(n)}}$, which can be seen as a KK excitation of the pseudo--Goldstone boson $G_{Z^{(0)}}$ associated with the $Z^{(0)}$ gauge boson. It is important to stress that the mass terms induced by compactification are invariant under the SGT. This is the reason why the KK effects decouple from SM observables in the heavy mass limit. The one--loop renormalizability of standard Green's functions is implicit in this. It is worth commenting that the couplings of the SM gauge fields to pairs of fermions $f^{(n)}f^{(n)}$ or $\hat{f}^{(n)}\hat{f}^{(n)}$ are vector--like, but the couplings involving flavor violating pairs $f^{(n)}\hat{f}^{(n)}$ are purely axial ($\gamma_\mu \gamma_5$). This is true for charged and neutral currents mediated by $W^{(0)\pm}_\mu$ and $Z^{(0)}_\mu$, respectively. Similarly, scalar and pseudoscalar neutral currents are mediated by the $H^{(0)}$ Higgs boson. However, charged currents mediated by KK excitations $W^{(n)\pm}_\mu$ have a general $V-A$ structure. A similar behavior is observed for neutral currents mediated by KK excitations $Z^{(n)}_\mu$ and $A^{(n)}_\mu$. Charged and neutral currents mediated by the $H^{(n)\pm}$, $H^{(n)}$, and $A^{(n)}$ scalar KK excitations also present a general scalar--pseudo scalar structure. Concerning the issue of quantization, a quantum action was defined by using the field--antifield formalism, in which the BRST symmetry arises naturally. Two gauge--fixing procedures that allowed us to remove the degeneration associated with the SGT and the NSGT were introduced. Since the SGT and the NSGT do not mix the zero modes and the KK excitations of the gauge parameters, such gauge--fixing procedures can be implemented independently of each other. So, for the SGT associated with the color group we introduced a conventional linear $R_\xi$--gauge, whereas in the case of the local NSGT, to which are subject the KK gluon excitations, we defined a $R_\xi$--gauge scheme that is covariant under the SGT of ${\rm SU_C}(3)$. As far as the gauge electroweak sector is concerned, the degeneration associated with the SGT of the electroweak group was removed by introducing a $R_\xi$--gauge procedure that is covariant under the electromagnetic group. On the other hand, to remove the degeneration associated with the NSGT we introduced gauge-fixing functions that transform covariantly under the SGT of the ${\rm SU_L}(2)\times {\rm U_Y}(1)$ group. The corresponding Faddeev--Popov ghost terms were expressed in terms of derivatives of the gauge--fixing functions, from which explicit Feynman rules can be derived.

\appendix
\renewcommand\thesection{Appendix \Alph{section}}
\section{The boson masses}
\label{bm}
 The mass term for the gauge fields is given by
\begin{eqnarray}
{\cal L}^{\rm gauge}_{\rm mass}&=&\sum_{n=0}\Bigg\{\left[\frac{1}{2}\left(\frac{n}{R} \right)^2+
\frac{g^2v^2}{8}\right]\left(W^{(n)1}_\mu W^{(n)1\mu}+W^{(n)2}_\mu W^{(n)2\mu} \right)
\nonumber \\&&
+\left(W^{(n)3}_\mu,B^{(n)}_\mu\right)M^{(n)}
\left(\begin{array}{ccc}
W^{(n)3\mu} \\
\, \\
B^{(n)\mu}
\end{array}\right)
\Bigg\} \, ,
\end{eqnarray}
where
\begin{equation}
M^{(n)}=\frac{1}{2} \left( \frac{n}{R} \right)^2I+M^{(0)} \, ,
\end{equation}
with $I$ representing the $2\times 2$ identity matrix and
\begin{equation}
M^{(0)}=\frac{v^2}{8}\left(\begin{array}{ccc}
\, \, g^2 & -gg' \\
\, \\
-gg' & \, \, g'^2
\end{array}\right)\,.
\end{equation}
The $M^{(n)}$ matrix is diagonalized by the well--known orthogonal matrix
\begin{equation}
R=\left(\begin{array}{ccc}
\, \, \, c_W & s_W \\
\, \\
-s_W & c_W
\end{array}\right)\, ,
\end{equation}
where $s_W$ and $c_W$ stand for the sine and cosine of the weak angle, given by $\tan \theta_W=g'/g$. The mass eigenstate fields and their corresponding masses are given by
\begin{eqnarray}
W^{(n)+}_\mu &=&\frac{1}{\sqrt{2}}\left(W^{(n)1}_\mu-iW^{(n)2}_\mu \right) \, , \, \, n=0,1, \cdots \, ,\\
W^{(n)-}_\mu &=&\frac{1}{\sqrt{2}}\left(W^{(n)1}_\mu+iW^{(n)2}_\mu \right) \, , \, \, n=0,1, \cdots \, , \\
m^2_{W^{(n)}}&=&\left(\frac{n}{R}\right)^2+m^2_{W^{(0)}}\, , \, \,  n=0,1, \cdots \, ,
\end{eqnarray}
\begin{eqnarray}
Z^{(n)}_\mu &=&c_W W^{(n)3}_\mu -s_WB^{(n)}_\mu \, , \, \, n=0,1, \cdots \, , \\
A^{(n)}_\mu &=&s_W W^{(n)3}_\mu +c_WB^{(n)}_\mu \, , \, \, n=0,1, \cdots \, , \\
m^2_{Z^{(n)}}&=& \left(\frac{n}{R} \right)^2+m^2_{Z^{(0)}}\, , \, \, n=0,1, \cdots \, , \\
m^2_{\gamma^{(n)}}&=&\left(\frac{n}{R} \right)^2\, , \, \,  \, n=0,1, \cdots \, ,
\end{eqnarray}
where $m_{W^{(0)}}$ and $m_{Z^{(0)}}$ are the SM masses for the $W$ and $Z$ gauge bosons, respectively.

Concerning the mass spectrum of the scalar fields, the up components of the Higgs doublets $\Phi^{(n)}$, denoted by $\phi^{(n)\pm}$, mix with the charged fields $W^{(n)\pm}_5\equiv \frac{1}{\sqrt{2}}\left(W^{(n)1}_5\mp iW^{(n)2}_5\right)$ as follows:
\begin{eqnarray}
{\cal L}^{\rm cs}_{\rm mass}&=&-\left(\phi^{(n)-},W^{(n)-}_5 \right)\left(\begin{array}{ccc}
\, \, \, \left(\frac{n}{R}\right)^2 & im_{W^{(0)}}\left( \frac{n}{R}\right)\\
\, \, \\
-im_{W^{(0)}}\left( \frac{n}{R}\right)& m^2_{W^{(0)}}
\end{array}\right) \left(\begin{array}{ccc}
\phi^{(n)+} \\
\, \, \\
W^{(n)+}_5
\end{array}\right) \nonumber \\
\,  \nonumber \\
&=&-m^2_{W^{(n)}}H^{(n)-}H^{(n)+}\, , \, \, \, n=1,2,\cdots \,,
\end{eqnarray}
where the physical fields $H^{(n)\pm}$ and the pseudo--Goldstone bosons $G^\pm_{W^{(n)}}$ are related to the original gauge eigenstates through the following unitary transformation
\begin{equation}
\left(\begin{array}{ccc}
H^{(n)+} \\
\, \\
G^+_{W^{(n)}}
\end{array}\right)=\left(\begin{array}{ccc}
c_\alpha & \, \, \, is_\alpha\\
\, \, \\
s_\alpha & -ic_\alpha
\end{array}\right)\left(\begin{array}{ccc}
\phi^{(n)+} \\
\, \\
W^{(n)+}_5
\end{array}\right)\, ,
\end{equation}
with the angle $\alpha$ given by
\begin{equation}
\tan\alpha=\frac{m_{W^{(0)}}}{\left( \frac{n}{R}\right)}\, .
\end{equation}
The pseudo--Goldstone bosons associated with the standard gauge fields $W^{(0)\pm}_\mu$ and $Z^{(0)}_\mu$ are, respectively, the up component of $\Phi^{(0)}$ and the imaginary part of the down component of $\Phi^{(0)}$:
\begin{equation}
\Phi^{(0)}=\left(\begin{array}{ccc}
G^+_{W^{(0)}} \\
\, \, \\
\frac{v+H^{(0)}+iG_{Z^{(0)}}}{\sqrt{2}}
\end{array}\right)\, ,
\end{equation}
where $H^{(0)}$ is the SM Higgs boson. On the other hand, the mass terms for the neutral fields $W^{(n)3}_5$, $B^{(n)}_5$, $H^{(n)}$, and $\phi^{(n)}_I$, being the latter two fields the real and imaginary parts of the down component of the $\Phi^{(n)}$ doublet, can be written as
\begin{eqnarray}
{\cal L}^{\rm ns}_{\rm mass}&=&-\frac{1}{2}m^2_{H^{(n)}}H^{(n)}H^{(n)}
\nonumber \\&&
-\frac{1}{2}\left(\phi^{(n)}_I,\hat{W}^{(n)3}_5\right)\left(\begin{array}{ccc}
\left(\frac{n}{R}\right)^2 & -m_{Z^{(0)}}\left(\frac{n}{R}\right)\\
\, \, \\
-m_{Z^{(0)}}\left(\frac{n}{R}\right) &  m^2_{Z^{(0)}}
\end{array}\right)\left(\begin{array}{ccc}
\phi^{(n)}_I \\
\, \\
\hat{W}^{(n)3}_5
\end{array}\right)
\end{eqnarray}
where we have carried out the following rotation
\begin{equation}
\left(\begin{array}{ccc}
W^{(n)3}_5 \\
\, \\
B^{(n)}_5
\end{array}\right)=\left(\begin{array}{ccc}
\, \, \, c_W & s_W\\
\, \, \\
-s_W &  c_W
\end{array}\right)\left(\begin{array}{ccc}
\hat{W}^{(n)3}_5 \\
\, \\
G_{A^{(n)}}
\end{array}\right)\, .
\end{equation}
 In the above expression, $G_{A^{(n)}}$ is the pseudo Goldstone boson associated with the gauge boson $A^{(n)}_\mu$. In addition, the masses of the KK excitations of $H^{(0)}$ are given by
\begin{equation}
m^2_{H^{(n)}}=\left(\frac{n}{R} \right)^2+m^2_{H^{(0)}}\, ,
\end{equation}
where $m_{H^{(0)}}$ is the SM Higgs mass. The above mass matrix can be diagonalized through the following orthogonal rotation

\begin{equation}
\left(\begin{array}{ccc}
\phi^{(n)}_I \\
\, \\
\hat{W}^{(n)3}_5
\end{array}\right)=\left(\begin{array}{ccc}
\, \, \, c_\beta & s_\beta \\
\, \, \\
-s_\beta &  c_\beta
\end{array}\right)\left(\begin{array}{ccc}
A^{(n)} \\
\, \\
G_{Z^{(n)}}
\end{array}\right) \, ,
\end{equation}
where $G_{Z^{(n)}}$ is the pseudo--Goldstone boson associated with the gauge KK mode $Z^{(n)}_\mu$ and $A^{(n)}$ represents a physical pseudoscalar field with mass given by $m_{A^{(n)}}=m_{Z^{(n)}}$ . In addition,
\begin{equation}
\tan \beta=\frac{m_{Z^{(0)}}}{\left(\frac{n}{R}\right)}\, .
\end{equation}
Notice that
\begin{equation}
\frac{\tan\alpha}{\tan\beta}=c_W \,.
\end{equation}

\section{Definitions in the Currents sector}
\label{A}
The covariant objects appearing in Eq. (\ref{currents}) are given by
\begin{eqnarray}
(D_\mu F)^{(0)}_L&=&D^{(0)}_\mu F^{(0)}_L
-\left(ig_s\frac{\lambda^a}{2}G^{(n)a}_\mu +ig\frac{\sigma^i}{2}W^{(n)i}_\mu +ig'\frac{Y}{2}B^{(n)}_\mu \right)F^{(n)}_L\, , \\ \nonumber \\
(D_\mu F)^{(n)}_L&=&D^{(0)}_\mu F^{(n)}_L-\left(ig_s\frac{\lambda^a}{2}G^{(n)a}_\mu+ig\frac{\sigma^i}{2}W^{(n)i}_\mu +ig'\frac{Y}{2}B^{(n)}_\mu \right)F^{(0)}_L \nonumber \\
&& -\Delta^{nrs}\left(ig_s\frac{\lambda^a}{2}G^{(r)a}_\mu+ig\frac{\sigma^i}{2}W^{(r)i}_\mu +ig'\frac{Y}{2}B^{(r)}_\mu \right)F^{(s)}_L \, ,\\ \nonumber \\
(D_\mu F)^{(n)}_R&=&D^{(0)}_\mu F^{(n)}_R
-\Delta'^{nrs}\left(ig_s\frac{\lambda^a}{2}G^{(r)a}_\mu+ig\frac{\sigma^i}{2}W^{(r)i}_\mu +ig'\frac{Y}{2}B^{(r)}_\mu \right)F^{(s)}_R \, ,
\end{eqnarray}

\begin{eqnarray}
(D_5F)^{(0)}_L&=&\left(ig_s\frac{\lambda^a}{2}G^{(n)a}_5+ig\frac{\sigma^i}{2}W^{(n)i}_5 +ig'\frac{Y}{2}B^{(n)}_5 \right)F^{(n)}_R\, ,\\ \nonumber \\
(D_5F)^{(n)}_L&=&\frac{n}{R}F^{(n)}_R
-\Delta'^{nrs}\left(ig_s\frac{\lambda^a}{2}G^{(r)a}_5+ig\frac{\sigma^i}{2}W^{(r)i}_5 +ig'\frac{Y}{2}B^{(r)}_5 \right)F^{(s)}_R \, , \\ \nonumber \\
(D_5F)^{(n)}_R&=&-\frac{n}{R}F^{(n)}_L-\left(ig_s\frac{\lambda^a}{2}G^{(n)a}_5+ig\frac{\sigma^i}{2}W^{(n)i}_5 +ig'\frac{Y}{2}B^{(n)}_5 \right)F^{(0)}_L\nonumber \\
&&-\Delta'^{nrs}\left(ig_s\frac{\lambda^a}{2}G^{(r)a}_5+ig\frac{\sigma^i}{2}W^{(r)i}_5 +ig'\frac{Y}{2}B^{(r)}_5\right)F^{(s)}_L \, ,
\end{eqnarray}

and

\begin{eqnarray}
(D_\mu \hat{f})^{(0)}_R&=&D^{(0)}_\mu f^{(0)}_R-\left(ig_s\frac{\lambda^a}{2}G^{(n)a}_\mu+ig'\frac{Y}{2}B^{(n)}_\mu \right)\hat{f}^{(n)}_R\, , \\ \nonumber \\
(D_\mu \hat{f})^{(n)}_R&=&D^{(0)}_\mu \hat{f}^{(n)}_R-\left(ig_s\frac{\lambda^a}{2}G^{(n)a}_\mu+ig'\frac{Y}{2}B^{(n)}_\mu \right)f^{(0)}_R
\nonumber \\&&
-\Delta^{nrs}\left(ig_s\frac{\lambda^a}{2}G^{(r)a}_\mu+ig'\frac{Y}{2}B^{(r)}_\mu \right)\hat{f}^{(s)}_R \, ,\\ \nonumber \\
(D_\mu \hat{f})^{(n)}_L&=&D^{(0)}_\mu \hat{f}^{(n)}_L -\Delta'^{nrs}\left(ig_s\frac{\lambda^a}{2}G^{(r)a}_\mu+ig'\frac{Y}{2}B^{(r)}_\mu \right)\hat{f}^{(s)}_L \, ,
\end{eqnarray}

\begin{eqnarray}
(D_5\hat{f})^{(0)}_R&=&\left(ig_s\frac{\lambda^a}{2}G^{(n)a}_5+ig'\frac{Y}{2}B^{(n)}_5 \right)\hat{f}^{(n)}_L\, ,\\ \nonumber \\
(D_5f)^{(n)}_R&=&\frac{n}{R}\hat{f}^{(n)}_L-\Delta'^{nrs}\left(ig_s\frac{\lambda^a}{2}G^{(r)a}_5+ig'\frac{Y}{2}B^{(r)}_5 \right)\hat{f}^{(s)}_L \, , \\ \nonumber \\
(D_5\hat{f})^{(n)}_L&=&-\frac{n}{R}\hat{f}^{(n)}_R-\left(ig_s\frac{\lambda^a}{2}G^{(n)a}_5+ig'\frac{Y}{2}B^{(n)}_5 \right)f^{(0)}_R
\nonumber \\&&
-\Delta'^{nrs}\left(ig_s\frac{\lambda^a}{2}G^{(r)a}_5+ig'\frac{Y}{2}B^{(r)}_5\right)\hat{f}^{(s)}_R \, .
\end{eqnarray}
In the above expressions, an appropriate application of the covariant derivative is assumed. For example, $D^{(0)}_\mu e^{(0)}_R=(\partial_\mu -ig'B^{(0)}_\mu Y/2 )e^{(0)}_R$, but $D^{(0)}_\mu d^{(0)}_R=(\partial_\mu-ig_sG^{(0)a}_\mu \lambda^a/2 -ig'B^{(0)}_\mu Y/2 )d^{(0)}_R$.

\section{The fermion masses}
\label{fm}
 As commented in Sec. II, the masses of the fermions emerge from both the Yukawa and the Currents sectors. The corresponding Lagrangian is given by
\begin{eqnarray}
-{\cal L}^{f}_{\rm mass}&=&\sum_{a,b=1}^3\Big\{\left(\lambda_{eab}\bar{L}^{(0)}_{La}e^{(0)}_{Rb}+\lambda_{dab}\bar{Q}^{(0)}_{La}d^{(0)}_{Rb}\right)\Phi^{(0)}_0
+\lambda_{uab}\bar{Q}^{(0)}_{La}u^{(0)}_{Rb}\tilde{\Phi}^{(0)}_0\nonumber \\
&&+\lambda_{eab}\left(\bar{L}^{(n)}_{La}\hat{e}^{(n)}_{Rb}+\bar{L}^{(n)}_{Ra}\hat{e}^{(n)}_{Lb} \right)\Phi^{(0)}_0
+\lambda_{dab}\left(\bar{Q}^{(n)}_{La}\hat{d}^{(n)}_{Rb}+\bar{Q}^{(n)}_{Ra}\hat{d}^{(n)}_{Lb} \right)\Phi^{(0)}_0 \nonumber \\ \nonumber \\
&&+\lambda_{uab}\left(\bar{Q}^{(n)}_{La}\hat{u}^{(n)}_{Rb}+\bar{Q}^{(n)}_{Ra}\hat{u}^{(n)}_{Lb} \right)\tilde{\Phi}^{(0)}_0\Big\}+\sum_{a=1}^3\Big\{\left(\frac{n}{R}\right)\bigg(\bar{L}^{(n)}_{aL} L^{(n)}_{aR}+\bar{\hat{e}}^{(n)}_{aR}\hat{e}^{(n)}_{aL}
\nonumber \\&&
+\bar{Q}^{(n)}_{aL}Q^{(n)}_{aR}+\bar{\hat{d}}^{(n)}_{aR}\hat{d}^{(n)}_{aL}
+\bar{\hat{u}}^{(n)}_{aR}\hat{u}^{(n)}_{aL}\bigg)\Big\}\, +\, {\rm H. \, c.}
\end{eqnarray}
In the flavor space, this Lagrangian can be written as follows:
\begin{eqnarray}
-{\cal L}^{f}_{\rm mass}&=&\bar{E}^{(0)}_L\Lambda_e E^{(0)}_R+\bar{E}^{(n)}_L\Lambda_e \hat{E}^{(n)}_R+\bar{E}^{(n)}_R\Lambda_e \hat{E}^{(n)}_L
\nonumber \\&&
+\left(\frac{n}{R}\right)\left(\bar{E}^{(n)}_LE^{(n)}_R +\bar{\hat{E}}\hspace{0.001cm}^{(n)}_R\hat{E}^{(n)}_L+\bar{N}^{(n)}_LN^{(n)}_R \right) \nonumber \\
&&+\bar{D}^{(0)}_L\Lambda_d D^{(0)}_R+\bar{D}^{(n)}_L\Lambda_d \hat{D}^{(n)}_R+\bar{D}^{(n)}_R\Lambda_d \hat{D}^{(n)}_L
\nonumber \\&&
+\left(\frac{n}{R}\right)\left(\bar{D}^{(n)}_LD^{(n)}_R +\bar{\hat{D}}\hspace{0.001cm}^{(n)}_R\hat{D}^{(n)}_L \right) \nonumber \\
&&+\bar{U}^{(0)}_L\Lambda_u U^{(0)}_R+\bar{U}^{(n)}_L\Lambda_u \hat{U}^{(n)}_R+\bar{U}^{(n)}_R\Lambda_u \hat{U}^{(n)}_L
\nonumber \\&&
+\left(\frac{n}{R}\right)\left(\bar{U}^{(n)}_LU^{(n)}_R +\bar{\hat{U}}\hspace{0.001cm}^{(n)}_R\hat{U}^{(n)}_L \right)+{\rm H.\, c.}\, ,
\end{eqnarray}
where $\Lambda_{e,d,u}=\frac{v}{\sqrt{2}}\lambda_{e,d,u}$ are matrices in the flavor space. Additionally,
\begin{equation}
E^{(n)}_{L,R}=\left(\begin{array}{ccc}
 e^{(n)}  \\
\, \, \\
\mu ^{(n)} \\
\, \\
\tau^{(n)}
\end{array}\right)_{L,R} \, , \, \, \, n=0,1,\cdots \, ; \, \, N^{(n)}_{L,R}=\left(\begin{array}{ccc}
 \nu_e^{(n)}  \\
\, \, \\
\nu_\mu ^{(n)} \\
\, \\
\nu_\tau^{(n)}
\end{array}\right)_{L,R} \, , \, \, \, n=1,2,\cdots \, ,
\end{equation}

\begin{equation}
D^{(n)}_{L,R}=\left(\begin{array}{ccc}
 d^{(n)}  \\
\, \, \\
s ^{(n)} \\
\, \\
b^{(n)}
\end{array}\right)_{L,R} \, , \, \, \, \,  U^{(n)}_{L,R}=\left(\begin{array}{ccc}
 u^{(n)}  \\
\, \, \\
c ^{(n)} \\
\, \\
t^{(n)}
\end{array}\right)_{L,R} \, ; \, \, \, \, n=0, 1,\cdots \, ,
\end{equation}

\begin{eqnarray}
\hat{E}^{(n)}_{L,R}=\left(\begin{array}{ccc}
 \hat{e}^{(n)}  \\
\, \, \\
\hat{\mu}^{(n)} \\
\, \\
\hat{\tau}^{(n)}
\end{array}\right)_{L,R} \, ,& \, \, \,
\hat{D}^{(n)}_{L,R}=\left(\begin{array}{ccc}
 \hat{d}^{(n)}  \\
\, \, \\
\hat{s} ^{(n)} \\
\, \\
\hat{b}^{(n)}
\end{array}\right)_{L,R} \, ,& \, \, \, \,  \hat{U}^{(n)}_{L,R}=\left(\begin{array}{ccc}
 \hat{u}^{(n)}  \\
\, \, \\
\hat{c} ^{(n)} \\
\, \\
\hat{t}^{(n)}
\end{array}\right)_{L,R} \, ;
 \, \, \, \, n= 1,2, \cdots \, ,
\end{eqnarray}

The mass eigenstates of the zero modes are determined by means of the standard unitary transformations
\begin{eqnarray}
N'^{(0)}_{L}&=&V^e_{L}N^{(0)}_{L} \, , \\
E'^{(0)}_{L,R}&=&V^e_{L,R}E^{(0)}_{L,R} \, , \\
D'^{(0)}_{L,R}&=&V^d_{L,R}D^{(0)}_{L,R} \, , \\
U'^{(0)}_{L,R}&=&V^u_{L,R}U^{(0)}_{L,R} \, .
\end{eqnarray}
Notice that, as it is usual, the left--handed neutrinos are rotated in the same way than the left--handed charged leptons. Regarding the excited KK modes, we impose the following transformations
\begin{eqnarray}
E'^{(n)}_{L,R}&=&V^e_{L}E^{(n)}_{L,R} \, , \, \, \, \, \hat{E}'^{(n)}_{L,R}=V^e_{R}\hat{E}^{(n)}_{L,R} \, , \\
D'^{(n)}_{L,R}&=&V^d_{L}D^{(n)}_{L,R} \, , \, \, \, \, \hat{D}'^{(n)}_{L,R}=V^d_{R}\hat{D}^{(n)}_{L,R} \, , \\
U'^{(n)}_{L,R}&=&V^u_{L}U^{(n)}_{L,R} \, , \, \, \, \, \hat{U}'^{(n)}_{L,R}=V^u_{R}\hat{U}^{(n)}_{L,R} \, .
\end{eqnarray}
We also demand that the neutrino excitations transform as the corresponding charged lepton excitations:
\begin{equation}
\label{tn}
N'^{(n)}_{L,R}=V^e_{L}N^{(n)}_{L,R} \,.
\end{equation}
Once carried out these transformations, one obtains
\begin{equation}
-{\cal L}^{f}_{\rm mass}={\cal L}^e_{\rm mass}+{\cal L}^d_{\rm mass}+{\cal L}^u_{\rm mass} \, ,
\end{equation}
where
\begin{eqnarray}
{\cal L}^e_{\rm mass}&=&\bar{E}'^{(0)}_LM^{(0)}_e E'^{(0)}_R+\left(\frac{n}{R}\right)\bar{\nu}^{(n)}_L\nu^{(n)}_R
\nonumber \\&&
+\sum_{a=1}^3\,m_{e^{(n)}_a}\, \left(\bar{e}'^{(n)}_{aL}\, \, \, \bar{\hat{e}}'^{(n)}_{aL} \right)\left(\begin{array}{ccc}
 \cos\alpha^{(n)}_{e_a} & \sin\alpha^{(n)}_{e_a}  \\
\, \, \\
\sin\alpha^{(n)}_{e_a} & \cos\alpha^{(n)}_{e_a}
\end{array}\right) \left(\begin{array}{ccc}
 e'^{(n)}_{aR}  \\
\, \, \\
\hat{e}'^{(n)}_{aR}
\end{array}\right)
+ \, {\rm H. \, c.}
\end{eqnarray}

\begin{eqnarray}
{\cal L}^d_{\rm mass}&=&\bar{D}'^{(0)}_LM^{(0)}_d D'^{(0)}_R
\nonumber \\&
+&\sum_{a=1}^3\,m_{d^{(n)}_a}\,\left(\bar{d}'^{(n)}_{aL}\, \, \, \bar{\hat{d}}'^{(n)}_{aL} \right)\left(\begin{array}{ccc}
 \cos\alpha^{(n)}_{d_a} & \sin\alpha^{(n)}_{d_a}  \\
\, \, \\
\sin\alpha^{(n)}_{d_a} & \cos\alpha^{(n)}_{d_a}
\end{array}\right) \left(\begin{array}{ccc}
 d'^{(n)}_{aR}  \\
\, \, \\
\hat{d}'^{(n)}_{aR}
\end{array}\right)
+ \, {\rm H. \, c.}
\end{eqnarray}

\begin{eqnarray}
{\cal L}^u_{\rm mass}&=&\bar{U}'^{(0)}_LM^{(0)}_u U'^{(0)}_R
\nonumber \\&
+&\sum_{a=1}^3\,m_{u^{(n)}_a}\,\left(\bar{u}'^{(n)}_{aL}\, \, \, \bar{\hat{u}}'^{(n)}_{aL} \right)\left(\begin{array}{ccc}
 \cos\alpha^{(n)}_{u_a} & \sin\alpha^{(n)}_{u_a}  \\
\, \, \\
\sin\alpha^{(n)}_{u_a} & \cos\alpha^{(n)}_{u_a}
\end{array}\right) \left(\begin{array}{ccc}
 u'^{(n)}_{aR}  \\
\, \, \\
\hat{u}'^{(n)}_{aR}
\end{array}\right)
+ \,{\rm  H. \, c.}
\end{eqnarray}
In the above expressions, $M^{(0)}_{e}={\rm diag}(m_{e^{(0)}},m_{\mu^{(0)}}, m_{\tau^{(0)}})$, etc. In addition,
\begin{equation}
\tan\alpha^{(n)}_{f_a}=\frac{m_{f^{(0)}_a}}{\left(\frac{n}{R}\right)} \, ,
\end{equation}
where $f_a$ stands for a charged lepton or quark. The mass eigenstates associated with the KK modes, which we will denote by $\tilde{f}_a$ and $\tilde{\hat{f}}_a$, are given by the following unitary transformations
\begin{equation}
\left(\begin{array}{ccc}
 f'^{(n)}_{aL}  \\
\, \, \\
\hat{f}'^{(n)}_{aL}\\
\end{array}\right)=V_L\left(\begin{array}{ccc}
 \tilde{f}^{(n)}_{aL}  \\
\, \, \\
\tilde{\hat{f}}\hspace{0.001cm}^{(n)}_{aL}\\
\end{array}\right)\, , \, \, \, \, \, \left(\begin{array}{ccc}
 f'^{(n)}_{aR}  \\
\, \, \\
\hat{f}'^{(n)}_{aR}\\
\end{array}\right)=V_R\left(\begin{array}{ccc}
 \tilde{f}^{(n)}_{aR}  \\
\, \, \\
\tilde{\hat{f}}\hspace{0.001cm}^{(n)}_{aR}\\
\end{array}\right)\, ,
\end{equation}
where
\begin{equation}
V_L=\left(\begin{array}{ccc}
 \cos\frac{\alpha^{(n)}_{f_a}}{2} & \, \, \sin\frac{\alpha^{(n)}_{f_a}}{2} \\
\, \, \\
\sin\frac{\alpha^{(n)}_{f_a}}{2} & -\cos\frac{\alpha^{(n)}_{f_a}}{2} \\
\end{array}\right)\, , \, \, \, \, \, V_R=\left(\begin{array}{ccc}
 \cos\frac{\alpha^{(n)}_{f_a}}{2} & -\sin\frac{\alpha^{(n)}_{f_a}}{2} \\
\, \, \\
\sin\frac{\alpha^{(n)}_{f_a}}{2} & \, \, \cos\frac{\alpha^{(n)}_{f_a}}{2} \\
\end{array}\right)\, .
\end{equation}
The $\tilde{f}_a$ and $\tilde{\hat{f}}_a$ states are degenerate, with mass given by
\begin{equation}
m_{f^{(n)}_a}=\sqrt{\left(\frac{n}{R}\right)^2+m^2_{f^{(0)}_a}} \, .
\end{equation}


\section*{Acknowledgments}
We acknowledge financial support from CONACYT and SNI (M\' exico). J.J.T. also acknowledges support from VIEP-BUAP under grant DES-EXC-2011.


\begin{thebibliography}{99}
%
\bibitem{ED} I. Antoniadis, {\it A possible new dimension at a few TeV}, Phys. Lett. \textbf{B} 246, 377 (1990);  N. Arkani-Hamed, S. Dimopoulos, and G. R. Dvali, {\it The hierarchy problem and new dimensions at a millimeter}, Phys.Lett.  \textbf{B} 429, 263 (1998); I. Antoniadis, N. Arkani-Hamed, S. Dimopoulos, andG. R. Dvali, {\it New dimensions at a millimeter to a fermi and superstrings at a TeV}, Phys. Lett. \textbf{B} 436, 257 (1998).
%
%
\bibitem{NT} H. Novales--S\' anchez and J. J. Toscano, {\it Gauge invariance and quantization of Yang-Mills theories in extra dimensions}, Phys. Rev. D \textbf{82}, 116012 (2010).
%
\bibitem{NT2} H. Novales--S\' anchez and J. J. Toscano, {\it Integration of Kaluza-Klein modes in Yang-Mills theories}, Phys. Rev. D \textbf{84}, 076010 (2011), e--print: arXiv:1105.2765 [hep-ph].
%
\bibitem{PS} J. Papavassiliou and A. Santamaria, {\it Extra dimensions at the one loop level: $Z\to b\bar{b}$ and $B-\bar{B}$ mixing}, Phys. Rev. D \textbf{63}, 016002 (2000).
%
\bibitem{ACD}T. Appelquist, H-C. Cheng, and B. A. Dobrescu, {\it Bounds on universal extra dimensions}, Phys. Rev. D \textbf{64}, 035002 (2001); T. Appelquist and H-U. Yee, {\it Universal extra dimensions and the Higgs boson mass}, Phys. Rev. D \textbf{67}, 055002 (2003).
%
\bibitem{FMNRT} A. Flores--Tlalpa, J. Monta\~no, H. Novales--S\' anchez, F. Ram\'irez--Zavaleta, and J. J. Toscano, {\it One--loop effects of extra dimensions on the $WW\gamma$ and $WWZ$ vertices}, Phys. Rev. D \textbf{83}, 016011 (2011).
%
\bibitem{NT3} H. Novales--S\' ancez and J. J. Toscano, {\it About gauge invariance in compactified extra dimensions}, Phys. Rev. D \textbf{84}, 057901 (2011).
%

\bibitem{HT} K. Fujikawa, {\it $\xi$--Limiting Process in Spontaneously Broken Gauge Theories}, Phys. Rev. D \textbf{7}, 393 (1973). See also, C. G. Honorato and J. J. Toscano, {\it $U_e(1)$--covariant $R_\xi$ gauge for the two--Higgs doublet model}, Pramana \textbf{73}, 1023 (2009), and references therein.
%
\bibitem{BRST} C. Becchi, A Rouet, and R. Stora, {\it Renormalization of the abelian Higgs--Kibble model}, Commun. Math. Phys.
\textbf{42}, 127 (1975); {\it Renormalization of gauge theories}, Ann. Phys. (N.Y.) \textbf{98}, 287 (1976); I.V. Tyutin, {\it Gauge Invariance in Field Theory and Statistical Physics in Operator Formalism}, FIAN (P.N: Lebedev Physical Institute of the USSR Academy of Science), Report No. 39, 1975.
%
\bibitem{AFAB} For a review, see J. Gomis, J. Paris, and S. Samuel, {\it Antibracket, antifields and gauge--theory quantization }, Phys. Rep. \textbf{259}, 1 (1995).
%

\end{thebibliography}
\end{document}